\documentclass[11pt]{article}

\pdfoutput=1
\usepackage{listings,courier}
\usepackage{physics}
\usepackage{graphicx}
\usepackage{subcaption}
\usepackage[font=scriptsize]{caption}
\usepackage{bm,color}
\usepackage{systeme}
\usepackage[section]{placeins}
\usepackage{placeins}
\usepackage{upgreek}
\usepackage{textcomp, gensymb}
\usepackage{fancyhdr}
\usepackage{hyphenat}
\usepackage{hyperref}
\usepackage[nobiblatex]{xurl}
\usepackage{multirow}
\usepackage[version=4]{mhchem}
\usepackage{amsmath}
\numberwithin{equation}{subsection}
\usepackage{array,booktabs,ragged2e}
\newcolumntype{R}[1]{>{\RaggedLeft\arraybackslash}p{#1}}
\newcolumntype{P}[1]{>{\centering\arraybackslash}p{#1}}
\usepackage{wrapfig}

\newcommand{\rev}[1]{#1}
\newcommand{\revtwo}[1]{#1}

\begin{document}

\title{An open-source pipeline for solving continuous reaction-diffusion models in image-based geometries of porous media} 

\author
{Justina Stark,$^{1, 2, 3}$ Ivo F. Sbalzarini$^{1, 2, 3,4}$\\
\\
 \normalsize{$^{1}$Technische Universit\"{a}t Dresden, Faculty of Computer Science, Germany.}\\
 \normalsize{$^{2}$Max Planck Institute of Molecular Cell Biology and Genetics, Dresden, Germany.}\\
 \normalsize{$^{3}$Center for Systems Biology Dresden, Dresden, Germany.}\\
 \normalsize{$^{4}$Cluster of Excellence Physics of Life, Dresden, Germany.}\\
 \\
}

\date{07.08.2023}

\maketitle

\begin{abstract}
We present a versatile open-source pipeline for simulating inhomogeneous reaction-diffusion processes in highly resolved, image-based geometries of porous media with reactive boundaries. Resolving realistic pore-scale geometries in numerical models is challenging and computationally demanding, as the scale differences between the sizes of the interstitia and the whole system can lead to prohibitive memory requirements. The present pipeline combines a level-set method with geometry-adapted sparse block grids on GPUs to efficiently simulate reaction-diffusion processes in image-based geometries. We showcase the method by applying it to fertilizer diffusion in soil, heat transfer in porous ceramics, and determining effective diffusion coefficients and tortuosity. The present approach enables solving reaction-diffusion partial differential equations in real-world geometries applicable to porous media across fields such as engineering, environmental science, and biology.
\end{abstract}


\section{Introduction}\label{sec:introduction}

Reaction-diffusion partial differential equations (PDEs) describe transport and conversion processes in porous media across science and engineering~\cite{Wakao1962, Muller2013Nodal, Beaudoin2018a, Cotterell2015}. Examples include soil contamination in geology~\cite{grathwohldiffusion}, drug formulations using amorphous solid dispersion in pharmacy~\cite{Alhijjaj2017}, transport of reactants in packed bed catalysts~\cite{Wakao1962, Haussener2009}, and biological phenomena such as tissue patterning and morphogenesis~\cite{Turing:1952vn, Wolpert1969, Crick1970, Wartlick2009, Kicheva2012, Muller2013Nodal, Yu2009}. The porous media in which these processes occur often have intricate, irregular geometries. 

These intricate geometries make constructing accurate numerical models of transport processes in porous media challenging. To simplify the problem, transport processes are often modeled by homogenizing the geometry~\cite{Ghanbarian2013, Valdes-Parada2017, Battiato2019, Hrouda2021, Bourbatache2022} and using coarse-grained geometric factors like tortuosity~\cite{Ray2018} and porosity. However, determining such factors is a non-trivial task, and simplified homogenized models often lack in accurately predicting \rev{reactive} transport processes when geometric heterogeneity is lost~\cite{Li2006}. To numerically model transport processes in porous media more accurately, or to computationally determine coarse-grained observables, the real-world geometry of the pores has to be factored in.  

Realistically modeling pores in a simulation, however, is challenging. One difficulty is that the surface of real-world irregular porous media geometries cannot be described by a global parametrization. Instead, an algorithmic surface representation has to be constructed that can be either explicit, as, e.g., surface triangulation~\cite{Boissonnat1984}, or implicit, as, e.g., phase fields~\cite{vanderWaals1893, Cahn1958} or level sets~\cite{Osher1988}.

Another challenge in resolving real-world pore geometries in a computer simulation is the difference in the length scales involved. While the interstitial space is often on the order of micrometers, the overall system size can be several orders of magnitude larger. A key challenge with such multiscale simulations is that they usually demand a lot of computer memory, limiting their feasibility and efficiency.

This memory requirement can be reduced in porous media simulations by taking advantage of the fact that only a fraction of the 3D space is of interest, and using a dense discretization is wasteful. Several methods allow for the placement of discretization points only in regions of interest. A famous example is particle methods, where the discretization points are not bound to any grid structure~\cite{Leonard1980, degond1989weighted, Pahlke2023}. The downside of lacking a grid structure, however, is that neighborhood information is not readily available, requiring additional data structures like cell lists~\cite{tildesley1987computer} or Verlet lists~\cite{Verlet1967}. Moreover, approximation of spatial derivatives requires computationally expensive mesh-free methods like \mbox{DC-PSE}~\cite{Schrader2010dcpse}. Hence, particle methods reduce the memory requirement thanks to their geometrical adaptivity but increase the computational cost due to their mesh-free structure.

A data structure that combines the geometrical adaptivity of particle methods with the structured property of a grid is the sparse block grid~\cite{Incardona2021}. Similar to particle methods and in contrast to dense grids, sparse grids enable the allocation of individual points in any subspace of the domain as needed. In contrast to particle methods, however, the points of a sparse grid are structured in Cartesian neighborhoods. Therefore, compared to particles, the overhead for storing positional information is smaller, neighborhood access becomes faster, and differential operators can be approximated using standard grid-based schemes. Furthermore, imposing boundary conditions on a sparse grid is computationally simpler than doing so on particles. Summarizing, owing to their grid-like structure and geometrical adaptivity, sparse block grids represent a suitable candidate for discretizing porous media geometries in a memory- and compute-efficient way.

Once the porous media geometry is numerically represented and discretized efficiently, it can be used to simulate reaction-diffusion processes. Reaction-diffusion processes can be described by PDEs, whose evolution in time requires the numerical approximation of differential operators. A wide range of methods for numerically approximating differential operators exists~\cite{Peiro2005}, including finite element methods~\cite{johnson2012numerical}, finite volume methods~\cite{moukalled2016finite}, finite difference methods~\cite{thomas2013numerical}, and spectral methods~\cite{gottlieb1977numerical}. Spectral methods cannot be adapted to sparse grids without losing their computational advantages. Finite element and finite volume methods can use adaptive, unstructured grids. Generation of an unstructured grid for a given complex geometry, however, has a high computational cost that might be justifiable when simulating fluid flow but is disproportionate when simulating reaction-diffusion processes. Thus, we use finite-difference methods and combine them with sparse grids to solve reaction-diffusion PDEs with high memory and computational efficiency and without the need for generating unstructured meshes.

To further accelerate simulations of transport processes in porous media, we take advantage of GPUs and parallelize our simulation using the Open Framework for Particles and Meshes (OpenFPM)~\cite{Incardona2019}. OpenFPM is well suited to our approach, as it provides a sparse-grid data structure for which tightly coupled scalability to multiple GPUs has been shown~\cite{Incardona2021}.

We extend the previous sparse block grid implementation by incorporating an implicit level-set algorithm for representing generic surfaces. This enables simulations in three-dimensional (3D), non-parametric, image-based geometries on distributed GPUs. The resulting open-source pipeline covers the entire workflow for porous media applications, from the segmentation of 3D sample images over the generation of a computational representation to solving the PDEs in the fully-resolved pore-scale geometries and visualizing the results.
As inputs, the present pipeline can handle different kinds of microscopy images or micro-computed tomography scans ($\upmu$CTs). From those, it generates realistic pore-scale geometries using the implemented level-set representation. On the process modeling side, we extend from homogeneous reaction-diffusion models to also include inhomogeneous reaction-diffusion processes, that is, to space-dependent diffusion coefficients and locally varying reaction rates. These can vary in function of the distance to the nearest interface, as that information is available from the level-set description of the geometry. We showcase the resulting image-based simulation pipeline in two real-world examples: inhomogeneous diffusion of fertilizer through soil and heat transfer in reticulate porous ceramics catalysts for renewable energy technology. We also benchmark the computational performance, multi-GPU scalability, and memory requirement of the present pipeline, comparing with dense-grid approaches.

\section{Methods}\label{sec:background}\label{sec:background:governingequations}

Numerical modeling of reaction-diffusion processes in porous media requires a mathematical description of the process, a numerical description of the geometry, and a suitable data structure to discretize space. We therefore first describe the governing equations and methods we use to build predictive models of reaction-diffusion processes in the image-based geometry of porous media. We also review the concepts of tortuosity and effective diffusivity.

Reaction-diffusion processes are modeled by a PDE describing the space-time evolution of a continuous concentration or density field. For isotropic bulk reaction-diffusion processes this PDE is:

\begin{equation}{\label{eq:background:governingequations:reaction_diffusion}}
    \pdv{u(\bm{x}, t)}{t} = \underbrace{\div (D(\bm{x}, t) \grad u(\bm{x}, t))}_{\text{diffusion}} + \underbrace{f(u(\bm{x}, t), \bm{x}, t)}_{\text{reactions}} \, ,
\end{equation}

where $u(\bm{x}, t)$ \rev{is a scalar field at time $t$ ($0 < t \leq t_{\mathrm{max}}$) and position $\bm{x}$ describing an intensive quantity, such as concentration or temperature. The diffusion domain $\Omega$ is} enclosed by a surface $\partial\Omega$, i.e., $\bm{x} \in \{\Omega \setminus \partial\Omega\}$. 
$D(\bm{x}, t)$ is the diffusion coefficient as a function of space and time, and $f(u(\bm{x}, t), \bm{x}, t)$ is the reaction term, \rev{which can be a reactive boundary, a linear combination of sources or sinks in the volume, both, or zero for pure diffusion.}

Diffusion through porous media is often inhomogeneous. For example, the diffusivity can vary between different phases or when molecules interact with the solid surface, leading to spatially varying coefficients~\cite{Elwinger2017}. The reaction term covers sources and sinks that can vary in space and time and depend on the current concentration via the law of mass action.

The solution of this PDE depends on the initial condition $u(\bm{x},0)$, the boundary conditions, and the geometry of the diffusion domain $\Omega$.

\subsection{Tortuosity and effective diffusivity}\label{sec:background:tortuosity}
The geometry of the diffusion space $\Omega$ affects the diffusive dispersion as it restricts the path a molecule can take~\cite{Ferreira2016, Sbalzarini2005frap}. Especially in the case of porous media with small interstitial space, a molecule has to move further on average to reach the same mean square displacement as in free space, because it has to circumvent obstacles~\cite{VanBrakel1974, Bohrer1984, Boudreau1996, Khirevich2011, Delgado2006, Shen2007}.

The ratio between the average path length $\langle L_\textrm{d}\rangle$ actually travelled by a diffusing molecule and the straight-line distance $L_\textrm{s}$ is called diffusive tortuosity, $\tau_\textrm{d}$~\cite{Smith1970, Ghanbarian2013}. The diffusive tortuosity $\tau_\textrm{d}$ can be expressed in terms of the diffusion coefficient in a bulk volume relative to its respective value in a tortuous medium~\cite{satterfield1963role}, as

\begin{equation}{\label{eq:background:tortuosity}}
    \tau_\textrm{d} = \left(\frac{\langle L_\textrm{d}\rangle}{L_\textrm{s}}\right)^2 = \frac{D_\textrm{f}}{D_{\mathrm{eff}}},
\end{equation}

where $D_\textrm{f}$ is the intrinsic molecular diffusion coefficient and $D_\textrm{eff}$ the effective diffusion coefficient in the tortuous medium~\cite{Shen2007, Ray2018}. The diffusive tortuosity or the effective diffusion coefficient of a material are of fundamental interest in engineering and science, because they can be used in macroscale models to make coarse-grained  predictions that help design and conceptualise diffusion in a material~\cite{Valdes-Parada2017, Ray2018}.

Many porous media, however, are heterogeneous, and averaging approaches do not capture local variations in tortuosity and the resulting anomalies~\cite{Ferreira2016, Garttner2020a}. Synthetic pore-scale models include tortuosity to generate periodic or stochastic networks of, e.g. cylinders or spheres. However, they do not reach the same level of heterogeneity as real-world geometries due to their artificial symmetry and lack of imperfect pore connectivity~\cite{Huang2019}.

\rev{Moreover}, pore geometry also affects chemical reactions. In particular, reactions at the interface depend on the surface area, which can differ significantly between media of different grain and pore sizes~\cite{Gray2018}. This is exploited when designing catalysts like packed beds and reticulate porous ceramics, which have a highly tortuous geometry with a large surface area~\cite{Haussener2009, Petrasch2008}.

Since characteristics like tortuosity, pore size, pore connectivity, and surface area significantly impact both diffusion and reactions, accurate predictive models of reaction-diffusion processes in porous media require taking into account the real pore-scale geometry of a given sample.

\subsection{Level-set method for image-based geometry reconstruction}\label{sec:background:level_set_method}
Real-world geometries for numerical models can be derived from 3D images, such as light-microscopy volumes or micro-computed tomography scans ($\upmu$CTs). For simulating transport processes in image-derived geometries, an algorithmic representation of the surface is required that forms the boundary for the transported medium.

Several methods to numerically represent surfaces exist. Explicit methods, like surface triangulation, directly discretize the surface by creating a surface grid~\cite{Boissonnat1984}. However, generating such unstructured grids has a computational cost disproportionate to the cost of the reaction-diffusion simulation~\cite{Friess2013}. Another disadvantage of these surface grids is that they have to be re-generated when the geometry deforms.

Alternatively, surfaces can be described implicitly by a higher-dimensional function that takes a specific value at the location of the surface. Implicit methods do not require an explicit discretization of the surface and therefore enable the representation of arbitrary geometries embedded in regular grids. Handling deformations also becomes more straightforward with implicit methods, as no surface grid needs to be moved.

The most prominent implicit surface methods are phase-field and level-set methods.  Phase-field methods belong to the class of diffuse-interface models with broad applications for simulating interface propagation in two-phase flows or mineral growth~\cite{Punke2022, han2022disconnection, Kelm2022}. Phase-field methods were first introduced by~\cite{vanderWaals1893} and later extended to volume-conserving phase-fields~\cite{Cahn1958}. Phase-field methods represent the different phases (e.g., solid vs.~fluid) of the material by a density function that varies smoothly around the interface.
A main limitation of phase-field methods is their high resolution requirement near the interface, where the phase function changes rapidly~\cite{Sethian2003}. Another issue is that the sharpness of the interface is limited by an arbitrarily chosen interface width parameter, leading to geometric inaccuracies and restrictions on the time-step size~\cite{Kim2000, Gibou2018}.

Level-set methods~\cite{Osher1988} overcome both problems by using a geometry-description function $\phi$ that has no additional physical meaning. The only purpose of the level function $\phi$ is to represent the interface as its zero iso-contour. A convenient choice for $\phi$ that is sufficiently flat around the surface, allows for continuous differentiation, and carries important geometric information is the signed-distance function SDF:
\begin{equation}\label{eq:level_set_method:SDF}
    \phi_\textrm{SDF}(\bm{x}) =
        \begin{cases}
        +d \quad & \text{for} \quad \bm{x} \in \Omega \\
        -d \quad & \text{for} \quad \bm{x} \in S\backslash\Omega \\
        \hfill 0 \quad & \text{for} \quad \bm{x} \in \partial\Omega.
    \end{cases}  
\end{equation}
Here, $d$ is the Euclidean distance from $\bm{x}$ to the closest point on the boundary $\partial\Omega$ of a domain $\Omega$ embedded in a space $S$. This implies that $|\grad\phi_\textrm{SDF}(\bm{x})|=1$ for all $\bm{x}$.

\begin{figure}[h]
    \centering
    \captionsetup{width=1.0 \textwidth}
    \includegraphics[width=1.0 \textwidth]{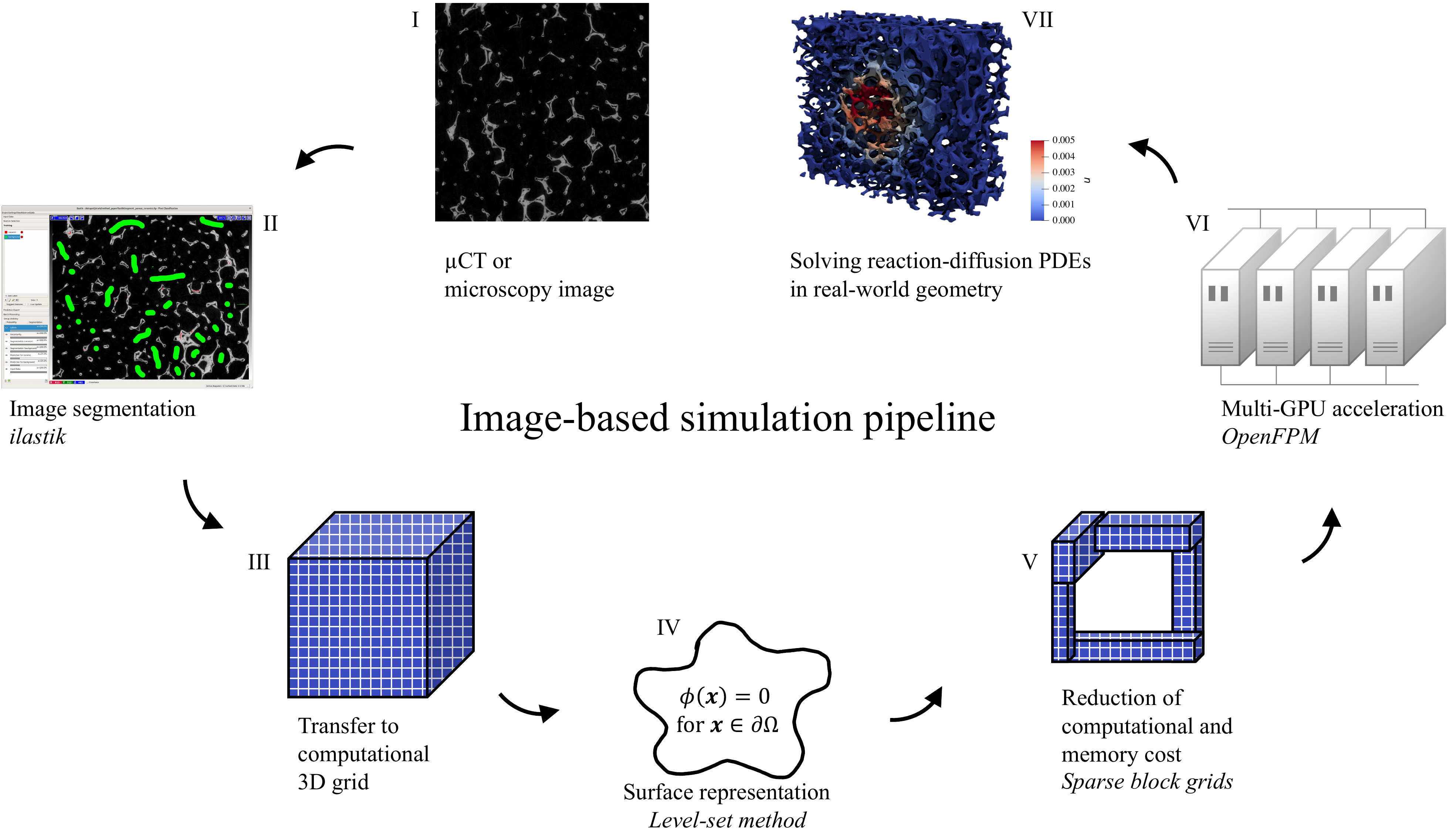}
    \caption{\rev{Overview of the present image-based simulation pipeline. The first step is the segmentation of a 3D $\upmu$CT or microscopy image (I) using the pixel-wise random-forest classifier (II) from ilastik~\cite{berg2019}. The binary segmentation mask is converted to an indicator function from which, on a dense grid (III), the level-set signed distance function (SDF) (IV) is obtained using Sussman redistancing. Based on the SDF, a sparse block grid~\cite{Incardona2021} geometry-adapted discretization is constructed (V). The sparse block grid data structure is then distributed over potentially multiple GPUs using the open-source scalable computing software OpenFPM~\cite{Incardona2019} (VI). Finally, simulation results are visualized using ParaView~\cite{ahrens2005} (VII).}}
    \label{fig:image_based_simulation_pipeline}
\end{figure}

The SDF is constructed for a geometry extracted from a 3D image. \rev{The entire workflow, as illustrated in Fig.~\ref{fig:image_based_simulation_pipeline}}, starts from binary segmentation of the image, classifying pixels into those that belong to $\Omega$ and those that do not. For pixel classification, we use the open-source software ilastik~\cite{berg2019}, which is based on trainable random-forest classifiers. The classification result is represented on the pixels by a binary indicator function. 
The SDF can be obtained from such an image-based indicator function by redistancing. A classic redistancing algorithm is the Sussman method~\cite{Sussman1999a}, which finds the SDF as the steady-state solution of the PDE
\begin{equation}\label{eq:level_set_method:sussman_redistancing}
    \pdv{\phi(\bm{x}, \alpha)}{\alpha} = \sigma(\phi(\bm{x},\alpha))(1-|\grad\phi(\bm{x},\alpha)|)
\end{equation}
for large artificial ``time'' $\alpha$. The smoothed sign function $\sigma$ proposed by~\cite{Peng1999a} is: 
\begin{equation}\label{eq:level_set_method:smooth_sign}
    \sigma(\phi(\bm{x}, \alpha)) = \frac{\phi(\bm{x}, \alpha)}{\sqrt{\phi(\bm{x}, \alpha)^2 + |\grad \phi(\bm{x}, \alpha)|^2 h^2}}\, ,
\end{equation}
where $h$ is the grid spacing, typically equal to the pixel size in the input image. 

Our implementation of Sussman redistancing for multi-CPUs on a dense grid is available as part of the open-source OpenFPM C++ library.
\rev{At the time of writing, our implementation computes $\grad\phi(\bm{x},\alpha)$ in Eq.~\eqref{eq:level_set_method:sussman_redistancing} using upwind finite differences of order 1, 3 (ENO), or 5 (WENO)~\cite{OsherFedkiw2003_book}. Since the level-set method imposes the limitation that neighboring interfaces must always be further apart than the width of the finite-difference stencil used for Sussman redistancing, we use first-order stencils for all results presented in this paper. They have the smallest width (namely $1h$) and therefore provide the highest spatial resolution for representing the geometry. We ensure a sufficiently large distance between interfaces by filtering geometric features of thickness $<2h$ from the binary indicator function before constructing the level-set.}

\rev{We also note that the SDF, while defining a mathematically sharp (i.e., not diffuse) interface, is not uniquely defined at sharp corners. During Sussman redistancing, rounding of corners with a radius of curvature $\approx h$ occurs. In image-derived geometries, however, this is not usually a concern, since the images rarely display sharp corners, due to the point-spread function of the imaging system. Moreover, the rounding error of level-set redistancing is of order $O(h)$.}
For these reasons, the error in the SDF converges with order one, as verified in Fig.~\ref{fig:appendix:figures:convergence_sussman}.

In our approach, we use the SDF not only to describe the interface geometry, but also to define consistent inhomogeneous diffusion coefficients, restrict reactions to the surface, and imposes no-flux Neumann boundary conditions for the diffusion. It can also be extended to describe deforming geometries, which we do, however, not consider here.

To obtain smoothly varying diffusion coefficients that depend on the distance to the surface, we make $D(\bm{x})$ a functional of $\phi_\text{SDF}(\bm{x})$, i.e.,

\begin{equation}\label{eq:background:inhomog_diffusion_coefficient}
    D_{\text{smooth}}(\phi_\text{SDF}(\bm{x})) = D_{\text{min}} + \frac{D_{\text{max}}}{1+\exp[-(\gamma_1+\gamma_2\cdot \phi_\text{SDF}(\bm{x}))]},
\end{equation}

in the case where the minimum and maximum diffusion coefficients ($D_{\text{min}}$, $D_{\text{max}}$) are known. The parameters $\gamma_1$ and $\gamma_2$ control the distance of the transition region from $\partial\Omega$ and the smoothing length, respectively. This transition region ensures that the diffusion coefficient can be continuously differentiated by finite difference~\cite{Fietier2013}.

For reactive transport processes that are restricted to one phase, it is no longer necessary to discretize the entire 3D domain $S$ once an implicit description of the diffusion space $\Omega$ has been obtained as a SDF. For example, we do not need grid nodes in the solid phase when simulating diffusion in the fluid phase. To ensure computational efficiency, we therefore aim to use a geometry-adaptive discretization method that does not waste memory and computation where they are not needed.

\subsection{Discretization and numerical methods}\label{sec:background:sparseGrids}
For multi-scale simulations in porous media, geometry-adaptive discretization data structures can significantly reduce memory requirements and computational costs. In particular, for processes that are restricted to one phase, using a dense grid would be wasteful. As mentioned in Section~\ref{sec:introduction}, particle methods are well suited for discretizing complex geometries but require computationally expensive data structures and methods to organize neighborhood information and approximate differential operators~\cite{Schrader2010dcpse, Pahlke2023}.

\begin{wrapfigure}{l}{0.3\textwidth}
    \centering
    \captionsetup{width=0.3\textwidth}
    \includegraphics[width=0.3\textwidth]{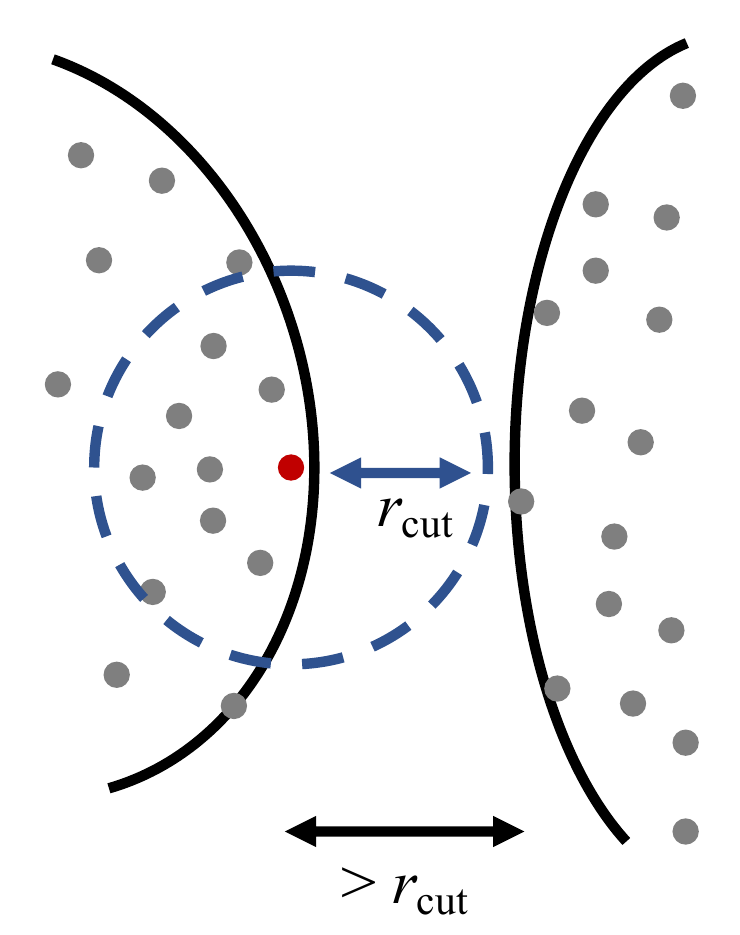}
    \caption{Illustration of the resolution requirement of particle methods in porous media geometries. Since operator kernels of particles in one interface must not overlap with particles on neighboring interfaces, a sufficiently dense particle distribution is required. This can be achieved by upsampling.}
    \label{fig:introduction:particle_resolution_requirement}
\end{wrapfigure}

The fact that operators are usually approximated over a neighborhood of particles within a specific kernel radius adds a further challenge for applications in porous media, where the interface surface is not singly connected. In particular, particle interaction neighborhoods must not overlap between different parts of the interface, as illustrated in Fig.~\ref{fig:introduction:particle_resolution_requirement}. This condition can only be guaranteed when pore-scale structures are simplified or there is more than one particle per pixel. In particular, for image-based geometries, the smallest distance between two interfaces is one pixel. Thus, there have to be sufficiently many particles per pixel to ensure that kernels will never overlap with particles from a neighboring interface. This limits the use of mesh-free particle methods for simulations in fully resolved porous media geometries.

A data structure that combines the advantages of a structured grid with the efficiency of geometry-adapted methods is the sparse block grid~\cite{Incardona2021}. It allows selectively allocating grid points in chunks to the phase of interest, thus considerably reducing the memory and computational overhead for discretizing sparse spaces. 

Yet, \rev{sparse grids preserve the advantage of a grid data structure of implying positional information and, thus, easily finding neighboring points.} Consequently, differential operators can be approximated using standard finite-difference schemes, which is faster than operator approximation on particles. Furthermore, imposing boundary conditions is simpler in finite-difference methods than in mesh-free particle methods. In particular, no mirror points in the other phase are required to impose no-flux Neumann boundary conditions. Instead, the no-flux condition is imposed within the finite-difference stencil. This is possible because the index of each neighbor within the stencil is known due to the grid structure, and the level-set function indicates on which side of the boundary a neighbor point lies. In the GPU kernel, the functor evaluating the finite difference is applied to all grid nodes. Within the functor, a stencil is constructed, and it is checked whether a neighbor is outside $\Omega$. If it is outside, its value is set equal to that of the center point so that the gradient between the center and that outside neighbor is zero.

\rev{For determining whether a point lies outside or inside $\Omega$, we store the level-set function on the sparse grid, along with the intensive scalar field $u$, the spatially varying diffusion coefficients, and the local reaction rates.} We numerically approximate the solution of Eq.~\eqref{eq:background:governingequations:reaction_diffusion} with the diffusion coefficient given by Eq.~\eqref{eq:background:inhomog_diffusion_coefficient} on the sparse grids using a Forward-Time Centered-Space (FTCS) finite-difference scheme of order two in space and order one in time. Although implicit time integrating schemes are more stable and allow for larger time step sizes, we use explicit time stepping as our focus is on GPU scalability and computational efficiency. The matrix inversions and global communication overhead incurred by implicit time stepping would quickly become prohibitive for large and complex-shaped domains. As our method addresses memory-limited problems, we chose the first-order explicit Euler time-stepping scheme because it does not require storing and communicating intermediate stages, as higher-order time-stepping schemes do. No-flux boundary conditions at the interface are imposed without a need for mirror points.

\rev{\subsection{Parallel implementation using OpenFPM}}

The present pipeline uses sparse block grids as implemented in the OpenFPM library for scalable scientific computing~\cite{Incardona2021}. \rev{OpenFPM is a C++ library for parallelizing particle- and mesh-based simulations~\cite{Incardona2019}. To benefit from the automatic parallelization capabilities of OpenFPM, we implement our simulation in C++ using the OpenFPM grid and sparse-grid data structures. OpenFPM generates these data structures at compile time using template meta-programming and thus hides memory access and communication management from the application, ensuring portability across CPU and GPU architectures~\cite{Incardona2023}.
This greatly simplifies the implementation of the present simulation, as demonstrated by the code examples in~\ref{sec:appendix:code_examples}. More code examples and a full API documentation are available on the OpenFPM website\footnote{At the time of writing: \url{http://openfpm.mpi-cbg.de/}}, including a documentation of our Sussman redistancing implementation and example functors for solving reaction-diffusion PDEs on sparse grids on the GPU.}

\rev{In the sparse-grid implementation of OpenFPM, we use a uniform chunk size of 8 grid points along each spatial dimension~\cite{Incardona2021}. These $8\times 8\times 8$ grid chunks are stored in an OpenFPM vector containing C++ aggregates. The chunk position in the dense grid is cached with an allocation mask for the grid points that are actually allocated in the sparse grid. Chunks are only stored if they contain at least one allocated grid point. Hence, the fewer the points needed to represent a geometry and the closer these points lie, the fewer chunks are required and, consequently, the larger the memory savings of a sparse grid compared to a dense grid implementation.}

In addition to reducing memory requirements, sparse grids also speed up the computation, as the finite-difference stencil only iterates over actually allocated points. \rev{The higher the density of points within an allocated chunk, the greater the thread efficiency on the GPU~\cite{Incardona2021}.}

\rev{The sparse grid can be distributed across multiple GPUs by domain decomposoition into cuboidal sub-grids. In OpenFPM, the domain decomposition is independent of the chunk boundaries of the sparse grid. Inter-GPU communication is done using CUDA-supported asynchronous MPI over the interconnect of the computer cluster~\cite{Incardona2021}.}

Since OpenFPM provides backends for both CUDA and HIP~\cite{Incardona2023}, our software implementation natively runs on GPUs from different vendors (including both Nvidia and AMD) and is scalable to distributed multi-GPU setups. \rev{Our implementation supports both 64-bit floating point (FP64) numbers and 32-bit floating point (FP32) numbers on the GPU. All results in this paper are obtained using FP32 precision, which is supported by all consumer and data-center GPUs and is typically faster~\cite{nvidia}.}

Together, these design choices significantly accelerate fully resolved pore-scale simulations in image-derived geometries of porous media, as we demonstrate below.\\

\section{Results}\label{sec:results}
We verify the correctness of our implementation of the methods explained in Sec.~\ref{sec:background}, showcase application examples to real-world reaction-diffusion problems in porous media, \rev{quantify the parallel scalability}, and compare the GPU performance and memory requirement of a sparse grid versus a dense grid. 

\subsection{Verification of the implementation}\label{sec:results:convergence}

\begin{figure}[h]
    \centering
    \includegraphics[width=1.0 \textwidth]{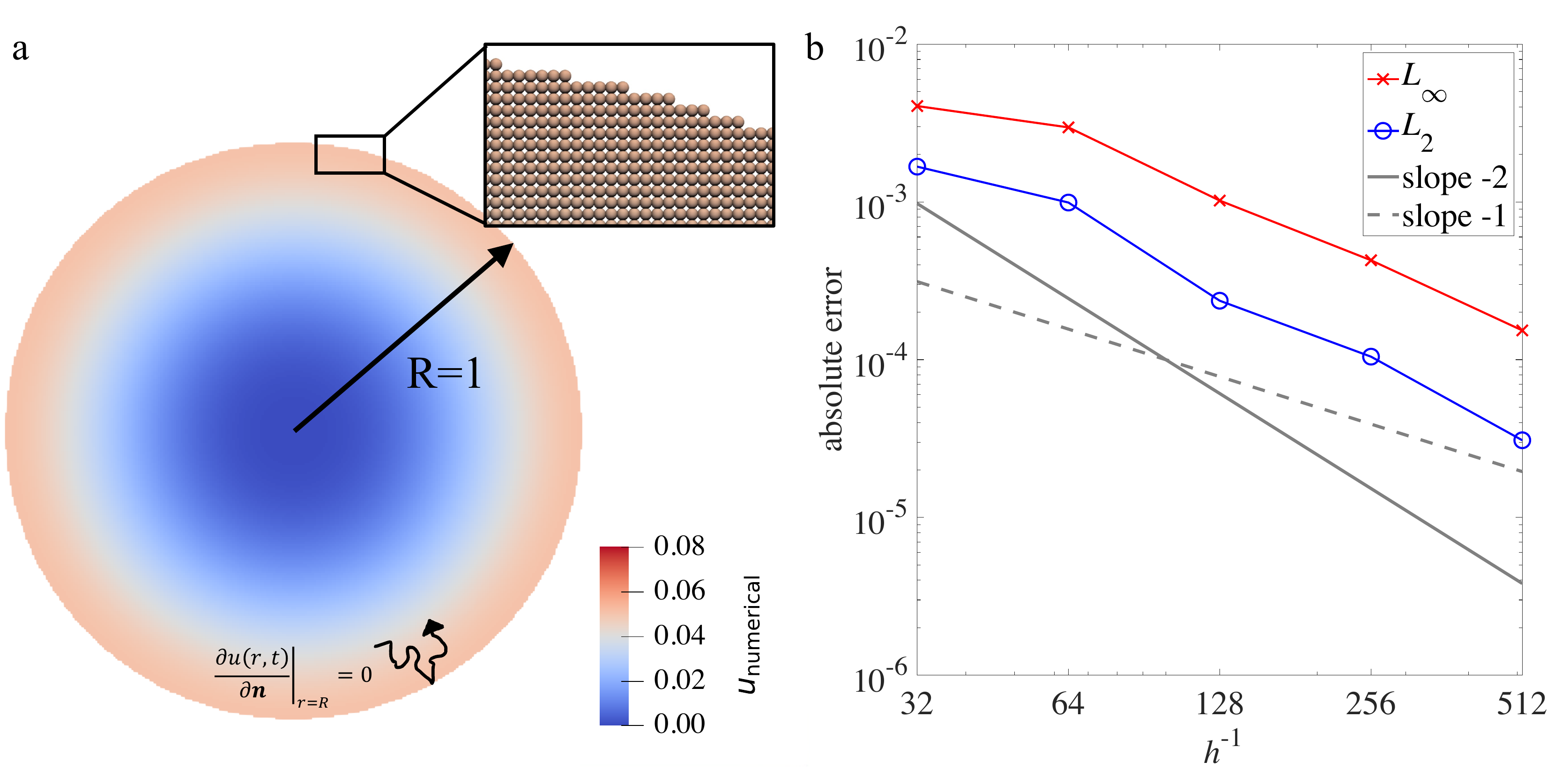}
    \caption{Verification of the implementation of diffusion on a 2D unit disk with no-flux Neumann boundary conditions. The simulation is run until a final time $t_{\rm final} = 0.025$ when the system is still far from steady state. The parameters are: $D=1.0$, $\Delta t=\frac{1}{8} h^2$. {\bf{(a)}} Visualization of the numerically computed concentration field at $t_{\rm final} = 0.025$ on a sparse grid (inset zoom). The sparse grid has been created based on the analytical signed-distance function of the unit disk. Time and diffusion coefficient values are in arbitrary units. {\bf{(b)}} Convergence of the absolute error of the overall method in $L_2$ (blue circles) and $L_\infty$ (red crosses) norms. 
    Grey lines show the expected slopes for first- (dashed) and second-order (solid) convergence.}
    \label{fig:results:convergence}
\end{figure}

To verify the accuracy of the method and the convergence of the discretization, we consider a benchmark case of homogeneous diffusion of a scalar field $u$ according to Eq.~\eqref{eq:background:governingequations:reaction_diffusion} \rev{without reactions, sources, or sinks, i.e., $f(u(\bm{x}, t), \bm{x}, t)\equiv 0$}. 
We make the test case as relevant as possible to the application of porous media, while still possessing an analytical solution against which convergence can be verified. Therefore, the test case includes a curved boundary embedded in a Cartesian sparse block grid, represented as a level set, with no-flux Neumann boundary conditions \rev{as described in Sec.~\ref{sec:background:sparseGrids}}. Specifically, we consider \rev{diffusion inside} a 2D disk of radius $R=1$, represented by a SDF and discretized using a sparse grid, as shown in Fig.~\ref{fig:results:convergence}a. 
\rev{The governing equation in polar coordinates ($r = \sqrt{x^2 + y^2}$ and $\theta = \atan(y, x)$) is:}
\begin{equation}{\label{eq:manufactured_solution:homogeneous_diffusion}}
    \pdv{u}{t} = 
    \pdv[2]{u}{r} 
    + \frac{1}{r} \pdv{u}{r} 
    + \frac{1}{r^2} \pdv[2]{u}{\theta}\, , 
\end{equation}
\rev{with boundary condition}
\begin{equation}{\label{eq:manufactured_solution:boundary_condition}}
    \left. \frac{\partial u(r, t)}{\partial \bm n} \right\rvert_{r=R=1} = 0
\end{equation}
\rev{and initial condition}
\begin{equation}{\label{eq:manufactured_solution:initial_condition}}
    u(r, 0) = \frac{r^3}{3} - \frac{r^4}{4}\, .
\end{equation}
\rev{The polar coordinates only serve to derive the exact analytical solution. The actual simulation is done in Cartesian coordinates in 2D.}
The analytical solution for this case, is derived in~\ref{sec:appendix:manufactured_solution} using the method of manufactured solutions~\cite{Roache2002}. All numerical solutions are compared against this exact solution in both the $L_2$ and $L_\infty$ norms of the absolute errors across all grid nodes, see Fig.~\ref{fig:results:convergence}b.

Like in the later applications, the time step size in the convergence study is adapted with the grid spacing. \rev{In particular, we fulfill the stability condition for diffusion in 2D,}
\begin{equation}{\label{eq:diffusion_stability_condition}}
    \Delta t < \frac{1}{2 \max D}  \frac{1}{\Delta x^{-2} + \Delta y^{-2}}\, ,
\end{equation}
\rev{by setting $\Delta t=\frac{1}{8} h^2$ for isotropic grid spacings $\Delta x = \Delta y = h$ and constant diffusion coefficient $D=1.0$}. 

For the smallest grid size of $32\times 32$ nodes, the disk's diameter is covered by approximately 16 grid nodes. The simulation is run until a final time of $t_{\rm final} = 0.025$ for all resolutions when the system is still far from steady state.

In order to separately quantify the error in solving the diffusion PDE in the domain and the error introduced by initial Sussman redistancing of the level set, we use the exact SDF for testing the PDE solver and, in addition, test the redistancing alone against the exact SDF. The convergence of the SDF error from redistancing alone is first-order, as shown in Fig.~\ref{fig:appendix:figures:convergence_sussman}.

For the finite-difference scheme used here, we expect first-order convergence in time and second-order convergence in space. \rev{The boundary conditions are imposed with first-order accuracy, as the curved boundary is approximated by the piecewise constant edge of the sparse grid, as illustrated in the inset of Fig.~\ref{fig:results:convergence}a. This is, however, not a limitation of the level-set method as such, which could in principle be used to approximate surface normals and curvature for higher-order boundary treatment. In the present workflow, however, we aim to benefit from GPU acceleration and sparse memory data structures, favoring a regular mesh. Moreover, since we use first-order Sussman redistancing, as described in Sec.~\ref{sec:background:level_set_method}, the error in the boundary conditions is not limiting. We therefore expect a convergence order between one, as dictated by the time stepping and Sussman redistancing, and two, as dictated by the spatial derivatives. This is confirmed in the convergence plot in Fig.~\ref{fig:results:convergence}b, where the empirical order of convergence is about 1.5, verifying the correctness of our implementation.}

\subsection{Generating real-world geometries from 3D images}\label{sec:application:image_based_reconstruction}

\begin{figure}[htbp!]
    \centering
    \includegraphics[width=1.0 \textwidth]{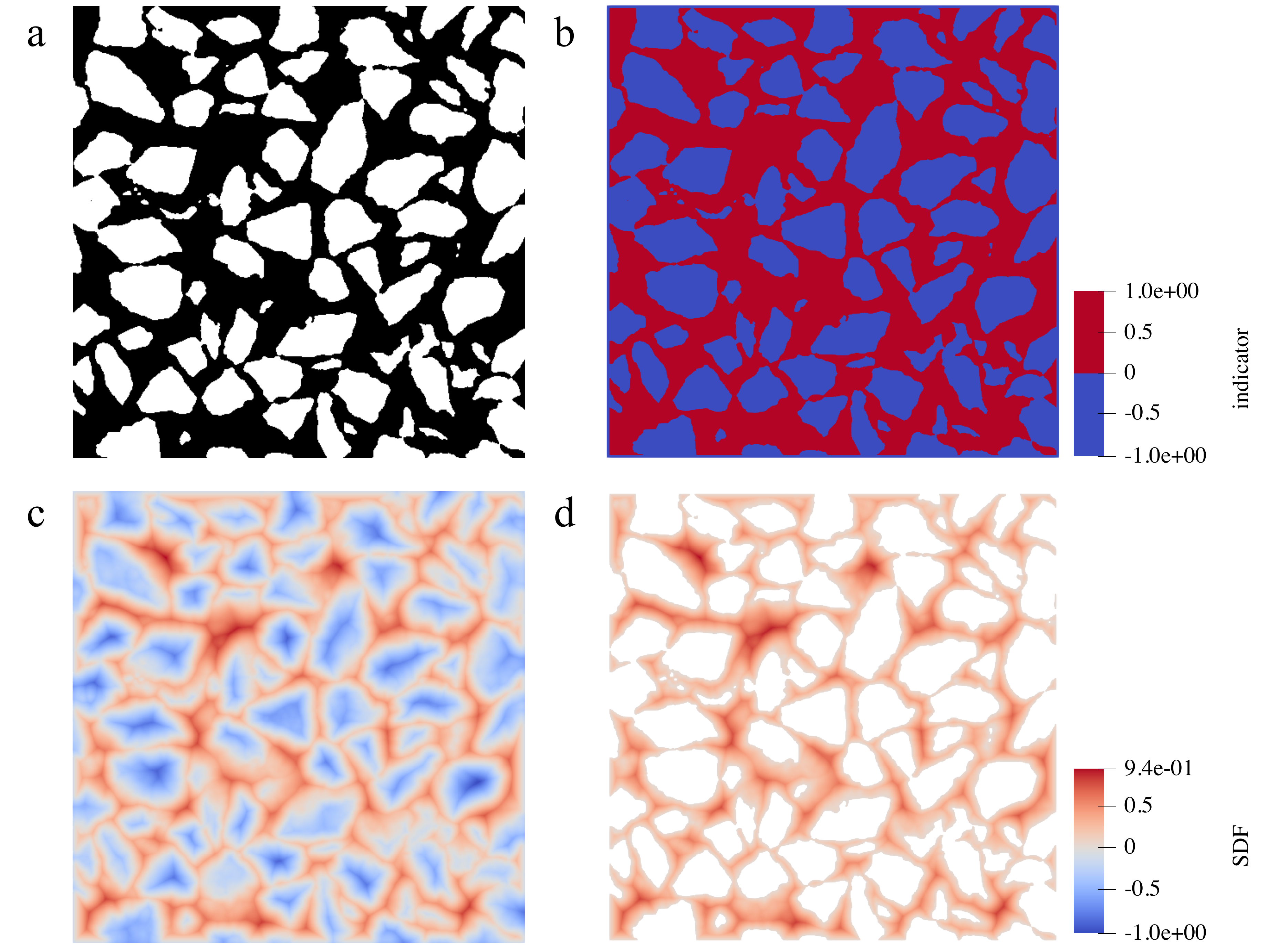}
    \caption{Steps of the present image-based geometry reconstruction pipeline, visualized for an exemplary slice through the 3D volume of a CaCO$_3$ packed bed~\cite{Haussener2009}. {\bf{(a)}} Segmented $\upmu$CT image showing the two phases with 45\,$\upmu$m voxel size~\cite{Haussener2009}. {\bf{(b)}} Image-based binary indicator function. {\bf{(c)}} Redistanced level-set signed-distance function (SDF) (Sec.~\ref{sec:background:level_set_method}). {\bf{(d)}} Sparse block grid containing only points in the diffusion phase, reducing storage.}
    \label{fig:applications:img_to_sparse_grid_2D}
\end{figure}

\begin{figure}[htbp!]
    \centering
    \captionsetup{width=1.0 \textwidth}
    \includegraphics[width=1.0 \textwidth]{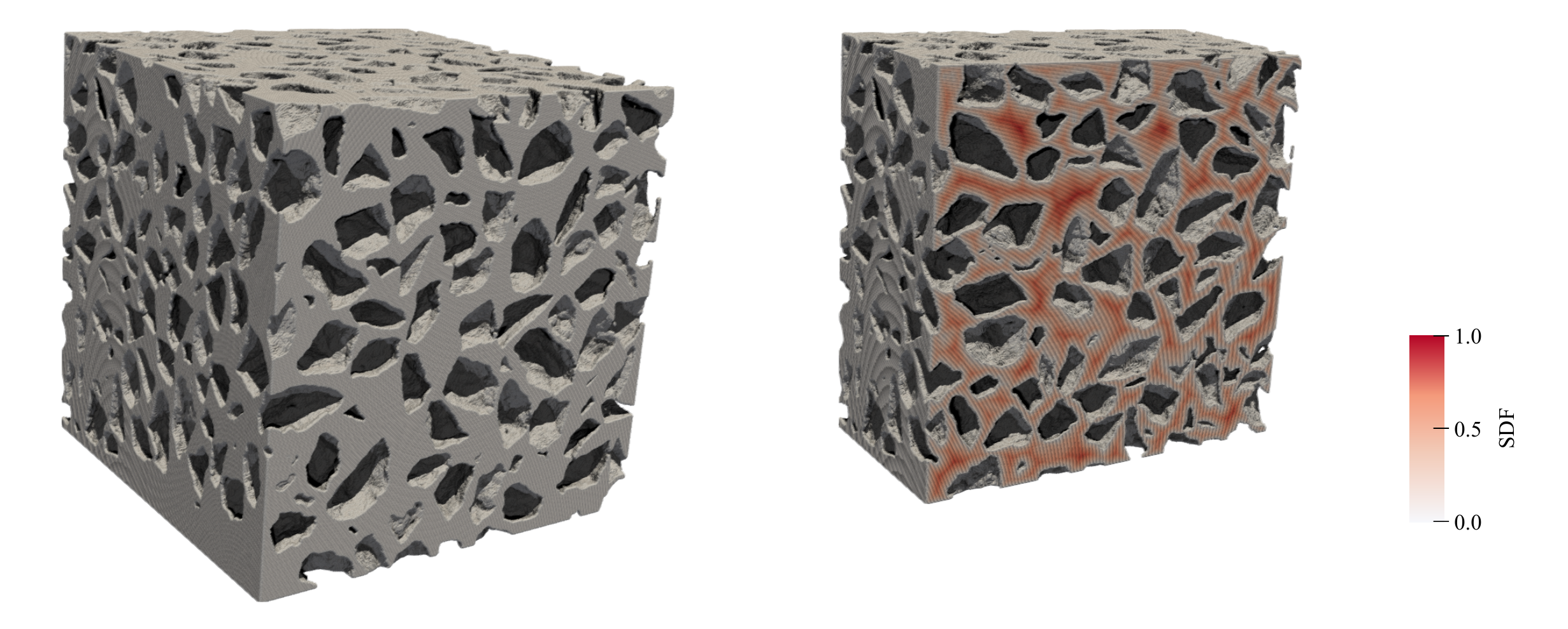}
    \caption{3D visualization of the sparse-grid level-set representation of the geometry from Fig.~\ref{fig:applications:img_to_sparse_grid_2D}d. Left: full block; Right: clipped block with SDF values overlaid in color.}
    \label{fig:applications:sdf_fluid_phase}
\end{figure}

We next illustrate the workflow of the present pipeline to go from a 3D image of a porous medium to a usable simulation domain. An overview of the full pipeline is provided in Fig.~\ref{fig:image_based_simulation_pipeline}. Fig.~\ref{fig:applications:img_to_sparse_grid_2D} shows the workflow for the case of a fluid-filled CaCO$_3$ packed bed. The segmented $\upmu$CT images (Fig.~\ref{fig:applications:img_to_sparse_grid_2D}a, voxel size 45\,$\upmu$m) were kindly provided by Prof.~J\"{o}rg Petrasch (Michigan State University, College of Engineering)~\cite{Haussener2009}. Our pipeline then first converts the binary segmentation to an indicator function $\{-1,1\}$ (Fig.~\ref{fig:applications:img_to_sparse_grid_2D}b). This is then used as an input to Sussman redistancing to generate the level-set SDF representation of the interface (Fig.~\ref{fig:applications:img_to_sparse_grid_2D}c) as described in Sec.~\ref{sec:background:level_set_method}. Based on this SDF, the sparse block grid is generated by only inserting points in the diffusion phase, e.g., in regions where $\phi_{\text{SDF}} > 0$. The SDF is then only stored restricted to those points, leading to a sparse geometry-adapted representation (Fig.~\ref{fig:applications:img_to_sparse_grid_2D}d). Allocation, however, is chunkwise meaning that a group of $8^{\rm{dims}}$ (with dimensions dims) points belonging to one chunk is allocated when a chunk contains at least one inserted point. A 3D visualization of the sparse-grid SDF is shown in Fig.~\ref{fig:applications:sdf_fluid_phase}.

\newpage
\subsection{Calculation of effective diffusion coefficient and tortuosity}\label{sec:applications:simulated_frap}

The effect of the pore geometry and topology on the diffusion dynamics is of interest in many applications~\cite{Satterfield1973, Bufe2017, Valdes-Parada2017} and is often studied in pore-scale numerical simulations~\cite{Yong2014, Rusinque2019, Ferguson2022}, as experimental determination is time-consuming and difficult~\cite{Boudreau1996}. The effect of spatial hindrance by a porous geometry on the diffusion of a molecule can be captured by an effective diffusion coefficient, which is related to the molecular diffusion coefficient by the diffusive tortuosity (Eq.~\eqref{eq:background:tortuosity}). While tortuosity causes an increase in diffusivity in advection-dominated systems, known as Taylor diffusion~\cite{Taylor1953}, the effective diffusivity of diffusion-dominated processes in tortuous geometries is smaller than the molecular diffusivity in free space~\cite{Akanni1987, Bohrer1984, Blum1989, Sbalzarini2005frap, Huang2019}. 

\rev{Over the past century, several theoretical and empirical correlations have been postulated that relate diffusive tortuosity $\tau_{\textrm{d}}$ to material porosity $\psi$. Many of them have the form}
\begin{equation}{\label{eq:tortuosity_porosity_relation}}
    \tau_\textrm{d} = \psi^{-N} \, ,
\end{equation}
\rev{with different exponents $N$ depending on \revtwo{the type of material~\cite{Bruggeman1935, Archie1942, Millington1959, Wakao1962}. For instance, it was shown that Eq.~\eqref{eq:tortuosity_porosity_relation} for $N=2$ fits experimental data of diffusion in low-density alumina and silver pellets, where diffusion is controlled by macropores~\cite{Wakao1962}. High-density pellets, where Knudsen diffusion and micropores dominate, however, could not be described by Eq.~\eqref{eq:tortuosity_porosity_relation}~\cite{Wakao1962}. This demonstrates that the quality of correlations like Eq.~\eqref{eq:tortuosity_porosity_relation} hinges on the similarity between the geometry in question and the geometry used during the fitting process. The CaCO$_3$ packed bed studied here has low density, with only $\approx$50\% of the sample belonging to the solid phase. We, therefore, expect the effective diffusion to be determined by the macropore tortuosity.} \revtwo{However, the porosity used in formulas like Eq.~\eqref{eq:tortuosity_porosity_relation}} is an average value over (some parts of) a material, which does not account for size limits of the interstitial space and for poor pore connectivity.

\revtwo{Tortuosity--porosity} relations have been comprehensively reviewed~\cite{Shen2007, Ghanbarian2013} and further refined by numerical simulations \cite{Wood2002, Ray2018, Huang2019}}. For granular media, like the CaCO$_3$ packed bed, Huang~\cite{Huang2019}, for example, found in random-walk particle simulations in synthetic geometries that the tortuosity can best be correlated with the porosity $\psi$ by the linear combination~\cite{Iversen1993}

\begin{equation}{\label{eq:applications:huang}}
    \tau_\textrm{d} = \psi + \beta (1 - \psi),
\end{equation}
with $\beta = 1.65$.

\begin{figure}[htbp!]
    \centering
    \captionsetup{width=1.0 \textwidth}
    \includegraphics[width=1.0 \textwidth]{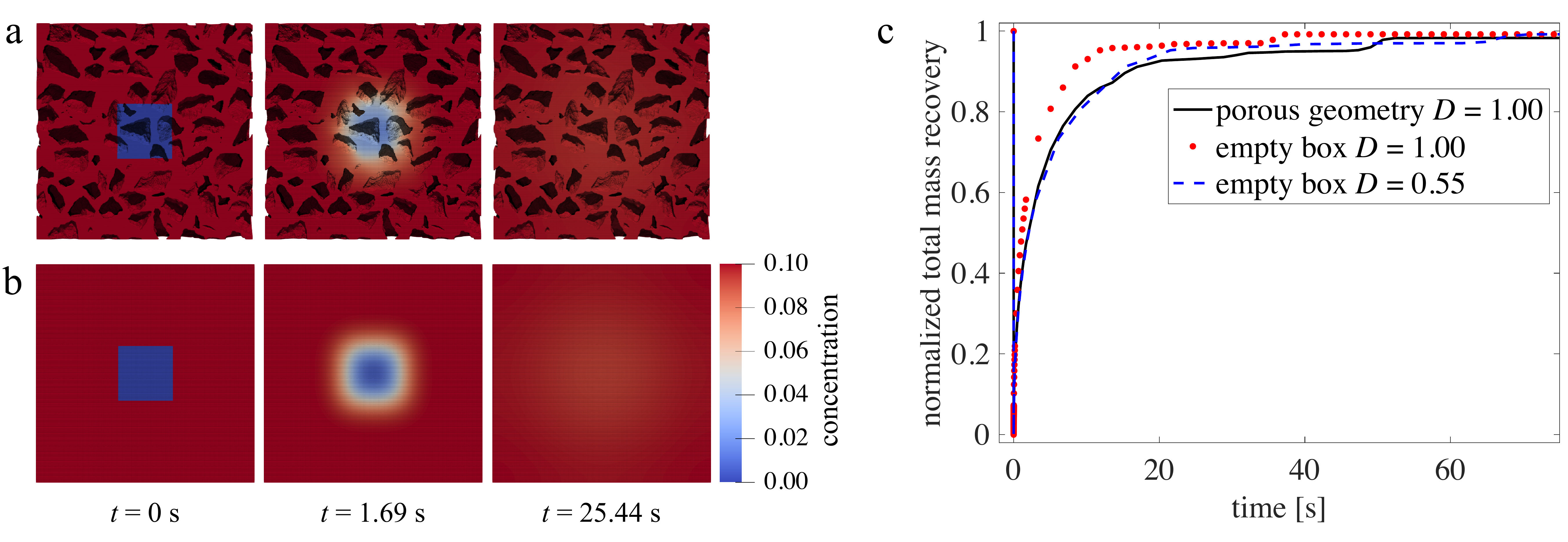}
    \caption{Simulation of fluorescence recovery after photobleaching (FRAP) in a CaCO$_3$ packed bed for determining the diffusive tortuosity and effective diffusivity using the present pipeline. {\bf{(a)}} Simulation snapshots of FRAP in the 3D image-based geometry of the porous medium with $D=1.0\,\text{mm}^2/\text{s}$. {\bf{(b)}} Simulation snapshot of FRAP in a bulk box of the same domain size with $D = 0.55\,\text{mm}^2/\text{s}$. To obtain the same recovery speed by free diffusion as in the porous structure, the diffusion coefficient had to be reduced by a factor of 1.82, thereby quantifying the diffusive tortuosity. {\bf{(c)}} FRAP curves showing the normalized mass in the initially bleached box for diffusion in the porous geometry with $D = 1.0\,\text{mm}^2/\text{s}$ and diffusion in the box with $D = 1.0\,\text{mm}^2/\text{s}$ and $D = 0.55\,\text{mm}^2/\text{s}$. The effective diffusion coefficient $D_{\rm{eff}} = 0.55\,\text{mm}^2/\text{s}$ was obtained by least-squares fit of the FRAP curve in the box to that in the porous medium.}
    \label{fig:applications:simulated_frap}
\end{figure}

In order to use the present pipeline to directly measure the tortuosity and effective diffusivity of a given porous geometry, we simulate fluorescence recovery after photobleaching (FRAP). FRAP is an experiment widely used in biology to determine effective diffusion coefficients in cells and tissues~\cite{Sbalzarini2005frap}. While FRAP experiments are not possible for inorganic solids, the simulation of FRAP in the respective image-derived geometry can still be done with the present pipeline and used to numerically determine the tortuosity.
We illustrate this by simulating FRAP in the image-based reconstruction of the fluid phase of the CaCO$_3$ from Section~\ref{sec:application:image_based_reconstruction} with $D=1.0\,\text{mm}^2/\text{s}$ (Fig.~\ref{fig:applications:simulated_frap}a). We then compare the so-simulated diffusion dynamics with diffusion in a bulk 3D box of the same edge length (Fig.~\ref{fig:applications:simulated_frap}b). 
Fitting (least squares) the free-space diffusion dynamics to those simulated in the porous medium, an effective diffusion coefficient of $0.55\ \mathrm{\mu m}^2/\mathrm{s}$ is found in the porous medium (Fig.~\ref{fig:applications:simulated_frap}c). According to Eq.~(\eqref{eq:background:tortuosity}), the diffusive tortuosity thus is:
\begin{equation}{\label{eq:applications:tortuosity}}
    \tau_\textrm{d} = \frac{D}{D_{\mathrm{eff}}} = \frac{1.0\,\text{mm}^2/\text{s}}{0.55\,\text{mm}^2/\text{s}} = 1.82.
\end{equation}
This means that diffusive transport is only about half as efficient in the fluid phase of a CaCO$_3$ than in a free-space bulk box.

This result agrees well with the value from Eq.~\eqref{eq:applications:huang}. In particular, for the sample porosity of $\psi=0.39$ as determined in Ref.~\cite{Haussener2009}, Eq.~\eqref{eq:applications:huang} predicts $\tau_d=1.40$, which is about 23\% lower than what we obtain with our FRAP simulation. The difference is in agreement with the finding in Ref.~\cite{Huang2019} that, due to unconnected pores, the real medium has a larger tortuosity than the synthetic geometries for which Eq.~\eqref{eq:applications:huang} was fitted. But even the tortuosity obtained by simulations in $\upmu$CT-based geometries most likely still underestimates the real hindrance to the molecular diffusion, as it ignores interactions with micropores that are not resolved in the $\upmu$CTs images~\cite{Elwinger2017}.

\subsection{Application example 1: Inhomogeneous diffusion of fertilizer in soil}\label{sec:applications:inhomogeneous_diffusion}

Understanding how solutes migrate through soil is of significant interest to several fields, including agriculture and environmental science~\cite{Shackelford1991, degryse2014, Essaid2015}. For example, fertilizer and pesticides used in agriculture are desired in specific regions where they protect and nourish the crops but undesired in other areas, where they pollute groundwater~\cite{Bijay-Singh1995, ElKhattabi2018} as a source of drinking water for many communities~\cite{Srivastav2020, Abd-Elaty2020}. Therefore, fertilizer application is ideally properly planned. To help farmers manage fertilizer application and prevent a buildup of excess chemicals in the soil, it is often a prerequisite to understanding diffusive transport in porous soil. This is our first application example where we showcase the use of the present computational pipeline.

When transport processes in soil are driven by steep concentration gradients, they are diffusion dominated, i.e., the P\'{e}clet number is small, and advection can be neglected. Modeling diffusion in the soil is not trivial, as the diffusive path of a molecule is not only hindered by macropore tortuosity of the porous soil structure, but in addition, the molecules can get trapped in micropores at the solid surfaces and chemically react with molecules in the solid phase~\cite{Elwinger2017, Rusinque2019}.

We therefore simulate fertilizer diffusion in soil by inhomogeneous diffusion of a concentration field in the fluid phase of a packed bed of CaCO$_3$ particles, using the image-based geometry reconstruction from  Fig.~\ref{fig:applications:sdf_fluid_phase}. In the simulated scenario, a fertilizer dissolved in the liquid enters from one face of the simulation domain and diffuses through the interstitial fluid phase between the solid CaCO$_3$ grains with an inhomogeneous diffusion coefficient that models diffusive hindrance by micropores close to the surface. In particular, $D_{\epsilon}(\bm{x}) = D \epsilon(\bm{x})$ with molecular diffusion coefficient $D=1.0\,\text{mm}^2/\text{s}$, microporosity $\epsilon = 1$ in the fluid, $\epsilon = 0$ in the solid. \rev{This follows the successful approach of locally modulating diffusion by a phase-dependent factor $\epsilon(\bm{x})$ to model boundary effects~\cite{Wakao1962, Soulaine2016, Etancelin2020, Hume2021}. Different from previous works, however, we introduce} a continuously differentiable transition zone between \rev{the phases}, which depends on the distance $\phi_\text{SDF}(\bm{x})$ to the fluid-solid interface according to Eq.~\eqref{eq:background:inhomog_diffusion_coefficient}. We here use smoothing parameters  $\gamma_1 = \gamma_2\phi_{\rm{min}}$, $\phi_{\rm{min}} = \min_{\bm{x} \in \Omega} \phi_{\text{SDF}}(\bm{x})$, and $\gamma_2 = 4h$. \rev{Also, in the present application, $\epsilon$ only represents microporosity at length-scales below the $\upmu$CT resolution, whereas the macroporosity $\psi$, resolved by the imaging, is explicitly modeled by the level-set function and the geometry-adapted sparse grid. Therefore, diffusion within macropores is simulated directly, while the hindrance by unresolved micropores at the solid surface is homogenized into an inhomogeneous diffusion constant, following the classic upscaling approach.} This fertilizer diffusion model can also readily be applied to contaminant diffusion.

\begin{figure}[htbp!]
    \centering
    \captionsetup{width=1.0 \textwidth}
    \includegraphics[width=1.0 \textwidth]{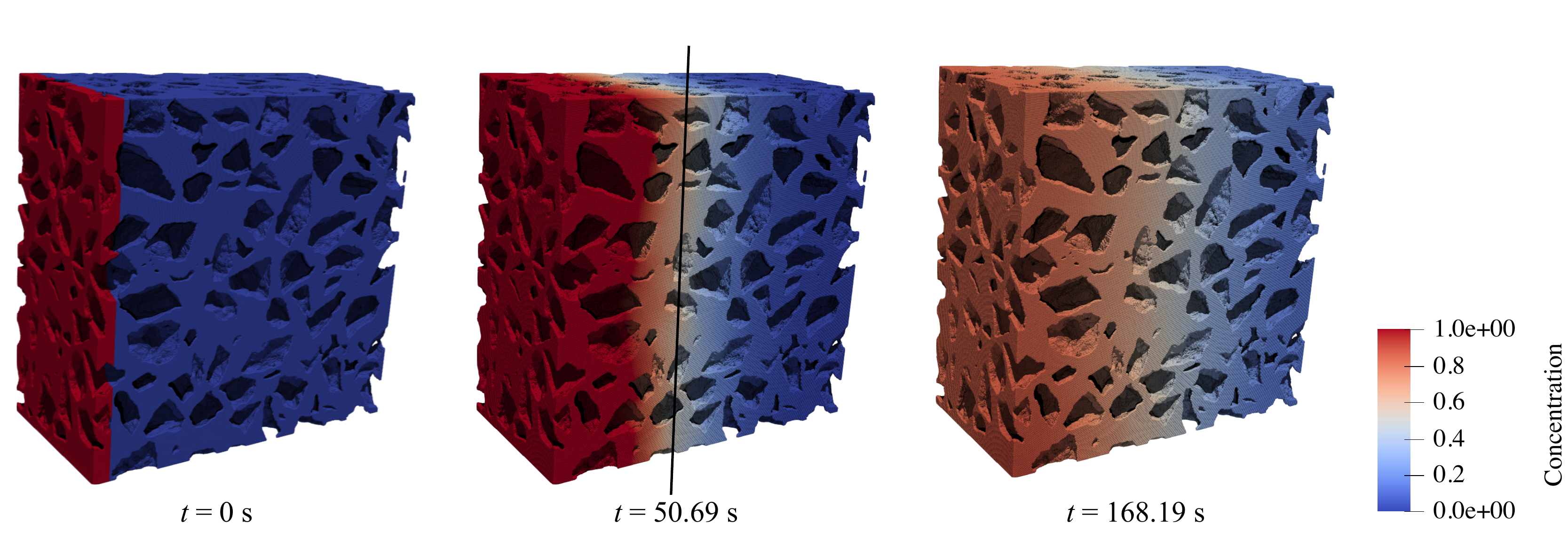}
    \caption{Simulation snapshots after 0, $4\times 10^5$, and $10^6$ time steps of a simulation of inhomogeneous diffusion of fertilizer in the fluid phase of a CaCO$_3$ packed bed. In the simulated scenario, the fertilizer dissolved in the fluid enters from one face of the simulation domain and diffuses through the fluid phase with an inhomogeneous diffusion coefficient that models hindrance by micropores at the CaCO$_3$ surfaces. This causes the diffusion front speed to depend on the locally varying pore geometry. The black vertical line indicates the location of a flat front with homogeneous speed for comparison.}
    \label{fig:applications:inhomogeneous_diffusion}
\end{figure}

Figure~\ref{fig:applications:inhomogeneous_diffusion} shows snapshots of the fertilizer diffusion simulation. It can be seen that the diffusion front becomes increasingly fuzzy over time. The increasing roughness of the diffusion front is caused by local variations in spatial hindrance and imperfect pore connectivity, i.e., the tortuosity of the porous geometry. Therefore, volume-averaged diffusion on the macroscale appears effectively anomalous. This anomalous behavior cannot be captured by geometry-averaging models with effective constants, demonstrating the importance of taking into account the pore-scale geometry.

\begin{wrapfigure}{r}{0.5 \textwidth}
    \includegraphics[width=0.5 \textwidth]{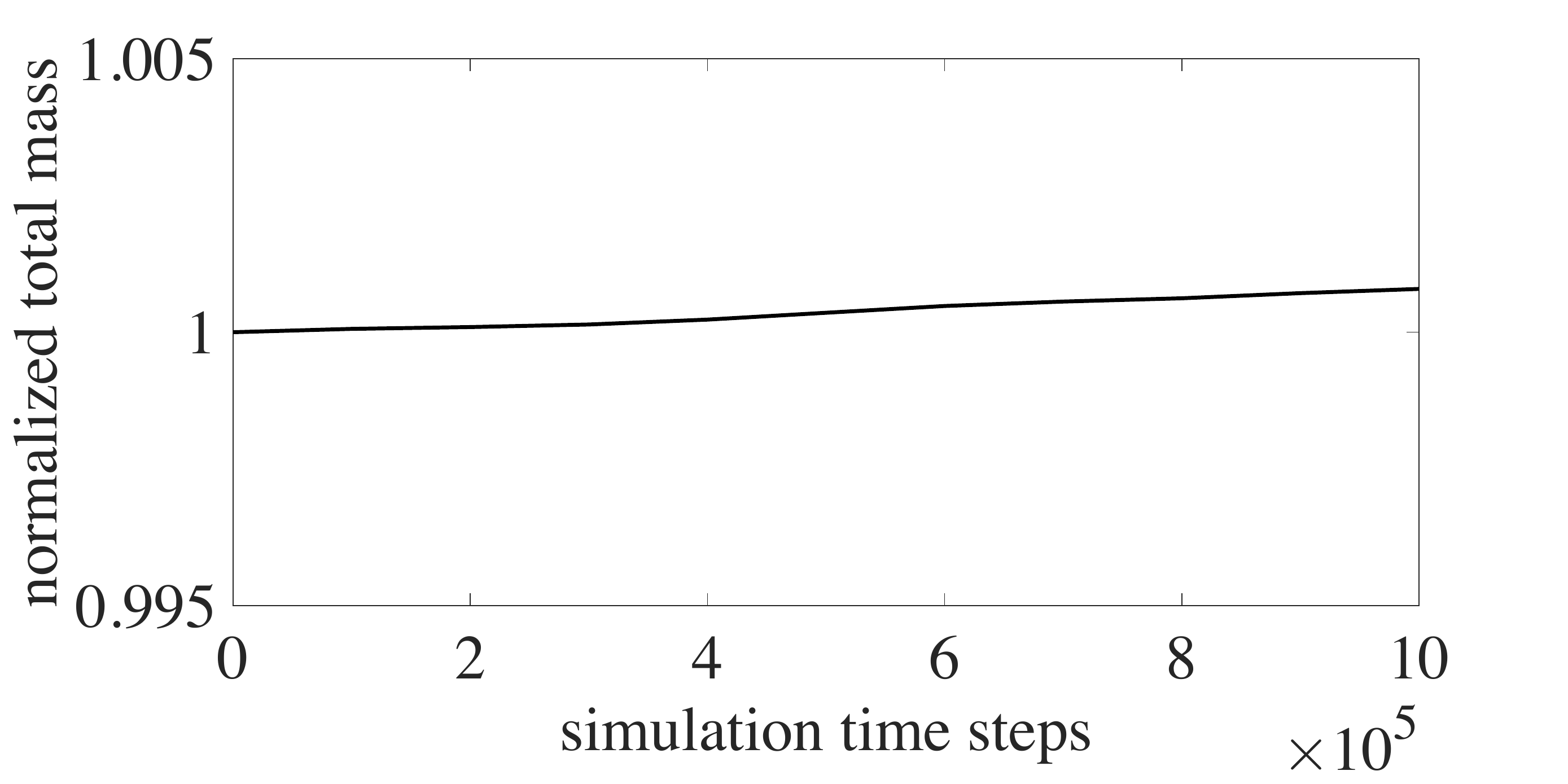}
    \caption{Mass conservation during inhomogeneous diffusion of fertilizer through the fluid phase of a CaCO$_3$ packed bed. The average increase in normalized total mass is $7.91 \times 10^{-10}$ per time step, thereby lying within the numerical round-off tolerance of the single-precision (FP32) arithmetics used.}
    \label{fig:applications:inhomogeneous_diffusion_total_mass}
\end{wrapfigure}

We test the accuracy of our implementation of the no-flux Neumann boundary conditions and the smoothed inhomogeneous diffusion coefficient by monitoring the conservation of mass during the simulation. The results in Fig.~\ref{fig:applications:inhomogeneous_diffusion_total_mass} show that the total mass is conserved to within the numerical round-off tolerance of the time-stepping scheme, as the simulation is done using single-precision arithmetics on the GPU. The average increase in normalized total mass per time step is $8 \times 10^{-10}$, and the total increase after $10^6$ time steps is $8 \times 10^{-4}$. 

We next compare the memory requirements and simulation GPU times of a dense-grid implementation with the sparse-grid version used in this application example. The results \rev{for different grid resolutions} are given in Table~\ref{tab:applications:comparison_dense_versus_sparse}. \rev{The dense grid is an OpenFPM block grid data structure with all chunks fully occupied.} Using the sparse block grid reduces the number of grid points in this particular example by 50\% from $150 \times 10^6$ points to $75 \times 10^6$ points. Memory requirements are reduced by 22\%, from 2.62\,GB to 2.04\,GB. This reduction is less than 50\% due to the additional bookkeeping data structures of the sparse block grid, and because a chunk of \rev{$8\times 8\times 8$} points is allocated whenever there is at least one point inside it, resulting in allocation of 77\% of all chunks in this example. 
The GPU time (excluding communication) and wall-clock time (including all necessary communication) per time step for the sparse block grid is around \rev{4.47\,ms} and \rev{4.48\,ms}, respectively, \rev{for the highest resolution} on a single Nvidia A100 GPU. This is \rev{13\%} faster than the dense grid implementation with \rev{5.11\,ms} and \rev{5.13\,ms}, respectively, on the same GPU. The speedup of a sparse over a dense grid in this example is modest because stencil operations are only performed for dense-grid points that lie in the fluid phase, requiring only one {\texttt{if}} condition more on a dense grid than on a sparse grid. Despite the CaCO$_3$ packed-bed sample being a rather dense geometry, however, the sparse grid offers some advantage, mainly in memory usage.

\subsection{Application example 2: Heat conduction through reticulate porous ceramics}\label{sec:applications:heat_diffusion_surface_reaction}

Another process that a reaction-diffusion PDE can describe is heat conduction through a solid. Heat conduction in porous media is relevant in many applications, ranging from thermal energy storage~\cite{Cao2013} to building insulation~\cite{Akolkar2017}. A class of materials with particularly beneficial characteristics for high-temperature applications are Reticulate Porous Ceramics (RPCs). RPCs play a role in applications including solar-thermal and thermo-chemical reactors for renewable energy technology~\cite{dhamrat2006numerical}, radiant burners~\cite{Howell1996}, and gas filters~\cite{Setten1999ceramic, Haussener2009}. We use this as a second application showcase for the present simulation pipeline.

\begin{figure}[htbp!]
    \centering
    \captionsetup{width=1.0 \textwidth}
    \includegraphics[width=1.0 \textwidth]{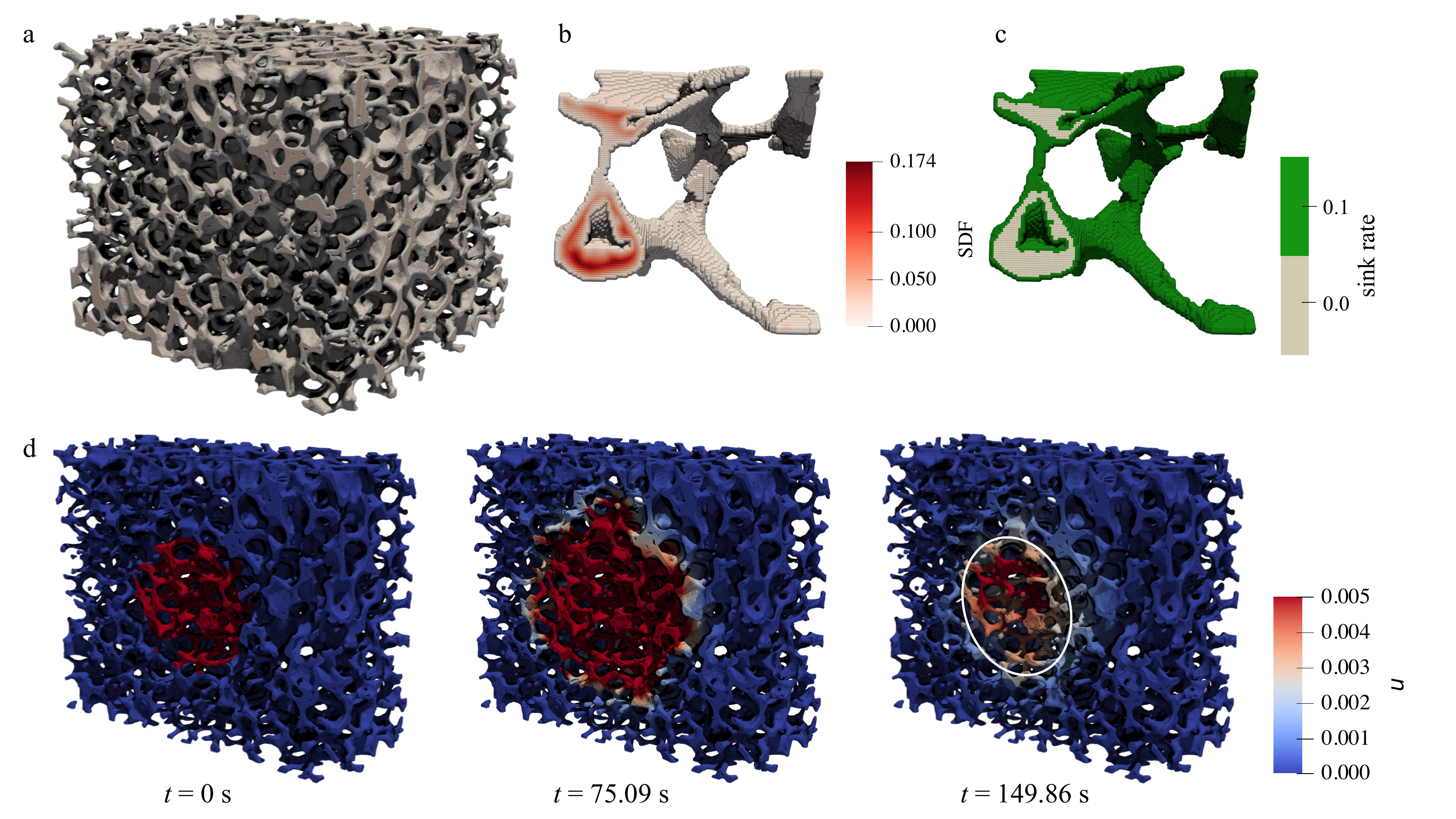}
    \caption{Diffusive heat conduction with distributed surface sink in a sample of reticulate porous ceramics. {\bf{(a)}} Surface rendering of the image-based reconstruction of the solid phase represented as a sparse-grid level set. {\bf{(b)}} Magnification of one exemplary pore with the level-set SDF shown in color to demonstrate the intricate shape of the pores. {\bf{(c)}} Heat-sink rate around the same pore to show how dissipation is restricted to the surface of the solid phase by the level-set function. {\bf{(d)}} Simulation snapshots shown as volume renderings in a clipped domain with temperature shown by color (arbitrary units). The initial condition is a spherical heat source at the center of the domain (left). From here, heat spreads through the solid with heat dissipation at the surface (middle, right). The diffusion front significantly deforms over time due to heterogeneous pore connections and the local variations in surface dissipation. The white circle indicates the location of a symmetric front with homogeneous speed for comparison.}
    \label{fig:applications:heat_diffusion_porous_ceramics}
\end{figure}

We therefore simulate diffusive heat conduction through the solid phase of a RPC sample with heat dissipation at the surfaces. The $\upmu$CT images of an RPC sample of size 1216$\times$1016$\times$941 voxels with a resolution of 30\,$\upmu$m/voxel were kindly provided by Prof.~J\"{o}rg Petrasch (Michigan State University, College of Engineering)~\cite{Petrasch2008, Haussener2010}. We segmented the images using the pixel-wise random-forest classifiers from ilastik~\cite{berg2019} and obtained the SDF of the solid phase using the approach described in Section~\ref{sec:application:image_based_reconstruction}. In the simulated scenario, heat diffuses through the RPC solid phase with a distributed heat sink at the solid surface. Using the level-set SDF, it becomes straightforward to constrain the heat sink to the surface as a reaction term (see Fig.~\ref{fig:applications:heat_diffusion_porous_ceramics}~b and~c).

Figure~\ref{fig:applications:heat_diffusion_porous_ceramics}~d shows snapshots of the simulation. The initial temperature distribution (left) is uniform, except for a sphere of radius $235 h$ in the center, where the temperature is higher. From here, temperature diffuses radially outwards with a diffusion coefficient of $D=0.1\,\text{mm}^2/\text{s}$. Heat is dissipated at the solid surface, which acts as a sink with rate $f(\bm{x})=0.1\,\text{s}^{-1}$ for $\bm{x} \in \partial\Omega$. Starting from a spherically symmetric initial condition, the temperature distribution becomes increasingly asymmetric and irregular over time (Fig.~\ref{fig:applications:heat_diffusion_porous_ceramics}d, middle), although a homogeneous diffusion coefficient is used. Like in Section~\ref{sec:applications:inhomogeneous_diffusion}, the diffusion front velocity is inhomogeneous (cf.~white circle in Fig.~\ref{fig:applications:heat_diffusion_porous_ceramics}d, right) due to the spatially varying tortuosity and heterogeneous pore connections in the RPC sample, as well as the local variations in surface area impacting heat flux due to the surface dissipation. This heterogeneity can only be captured when considering the pore-scale geometry. 

We again compare the computational cost of the present sparse-grid implementation with that of a \rev{dense block} grid in a domain of 1216$\times$1016$\times$941 grid cells. This shows a significant reduction in allocated memory and GPU time as summarized in Table~\ref{tab:applications:comparison_dense_versus_sparse}. This is because the RPC solid phase is truly sparse, filling only about 9\% of the space, with points spread over 21\% of the grid chunks. The number of mesh nodes in the sparse grid is thus reduced to $10^8$ from the $1.2 \times 10^9$ of a full grid, which is an 11-fold reduction. This significantly reduces the memory requirement from \rev{20.3}\,GB (dense) to 4.2\,GB (sparse).

Again, the reduction in memory requirement is less than in the number of grid points, due to the \rev{$8\times 8\times 8$} chunk size of the sparse block grid.
The GPU time (excluding communication) and wall-clock time (including all necessary communication) per time step for the sparse grid are around \rev{3.79\,ms} and \rev{4.12\,ms}, respectively, \rev{for the highest resolution} on two Nvidia A100 GPUs connected via NVLink. This is \rev{76\%} faster than the dense grid implementation with \rev{16.68\,ms} and \rev{16.97\,ms}, respectively, on the same two GPUs. Compared with the example from  Sec.~\ref{sec:applications:inhomogeneous_diffusion}, this demonstrates that a sparser geometry leads to better performance gains. 

\newpage
\rev{\subsection{Performance and scalability}\label{sec:applications:performance}}

\rev{We benchmark computational performance of our code for different grid resolutions and on different numbers of GPUs. All benchmarks are performed on A100-SXM4 GPUs of the \textit{taurus} computer of TU Dresden, which has 8 GPUs per compute node. GPUs within a node are connected by NVLink, whereas individual nodes are connected by 200\,Gb/s Infiniband. We use Nvidia's {\it{Unified Communication - X Framework}} (UCX), as integrated into OpenMPI, with the {\tt{SKIP\_LABELLING}} option enabled, since the grid nodes do not move.}

\begin{figure}[htbp!]
    \centering
    \captionsetup{width=1.0 \textwidth}
    \includegraphics[width=1.0 \textwidth]{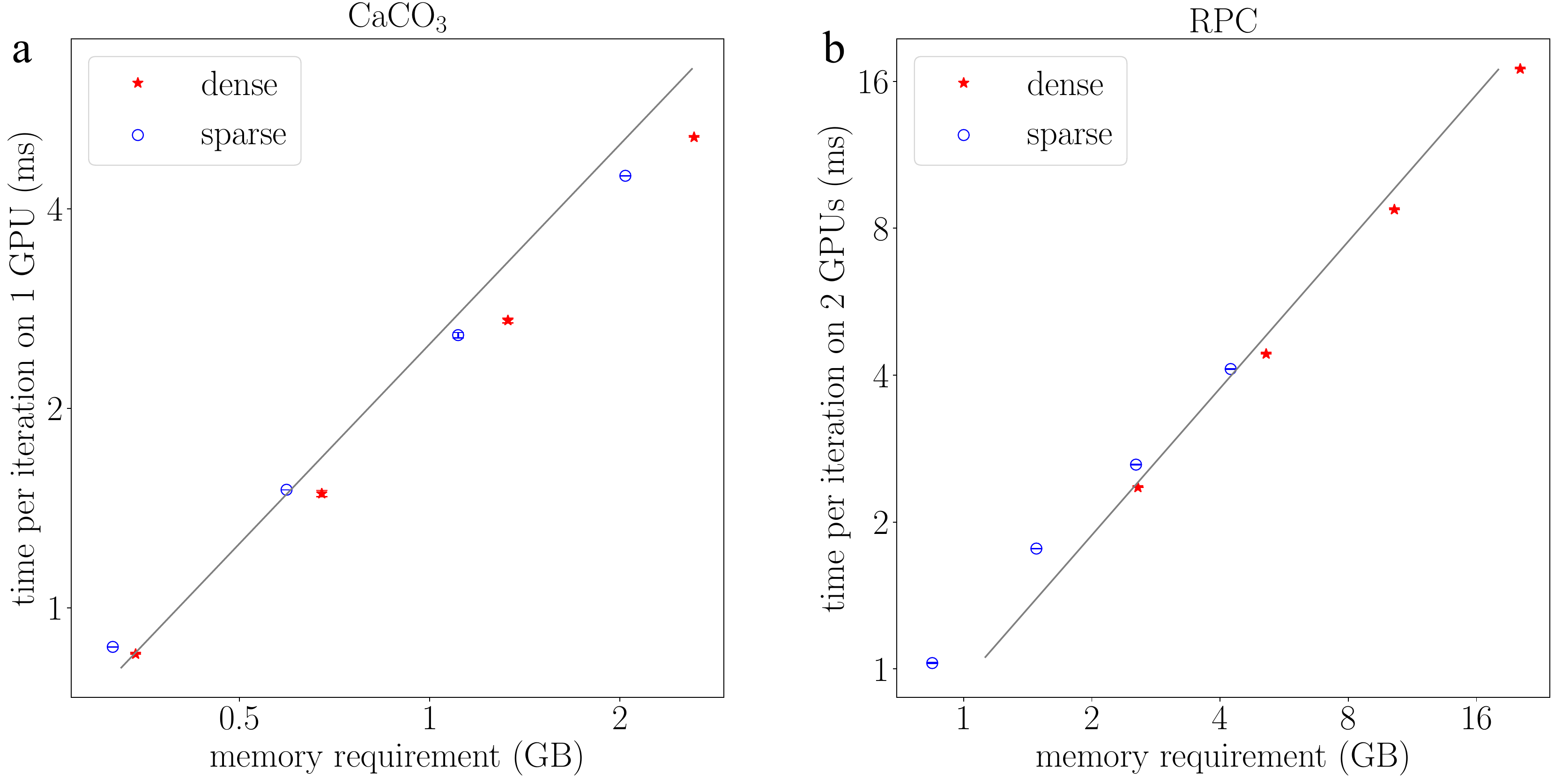}
    \caption{\rev{Plots of the data from Table \ref{tab:applications:comparison_dense_versus_sparse} comparing the wall-clock times per time step and the memory requirements for dense (red stars) and sparse (blue circles) block grids on a fixed number of GPUs. Error bars show the standard error; solid lines denote linear scaling. {\bf{(a)}} Inhomogeneous diffusion in a CaCO$_3$ particle-packed bed on a single Nvidia A100 GPU. {\bf{(b)}} Heat conduction in reticulate porous ceramics on 2 Nvidia A100 GPUs connected via NVLink.}}
    \label{fig:applications:problem_size_scaling}
\end{figure}

\rev{We first measure how the memory requirements and wall-clock times scale with problem size for both dense and sparse grids. The results are given in Table~\ref{tab:applications:comparison_dense_versus_sparse} and Fig.~\ref{fig:applications:problem_size_scaling}. In this benchmark, ``dense grid'' refers to a fully occupied (i.e., all chunks are allocated) OpenFPM block grid data structure. The grid resolution is successively halved along one dimension at a time, so the expected scaling is linear (solid line in Fig.~\ref{fig:applications:problem_size_scaling}). Although the fully occupied chunks of the dense grid enable better thread efficiency on the GPU than the partly filled chunks of the sparse grid, Fig.~\ref{fig:applications:problem_size_scaling} confirms that the sparse grid scales almost as well as the dense grid, and both scale almost linearly.}

\begin{figure}[h]
\footnotesize
\centering
\raisebox{-.5\height}{\includegraphics[width=0.45\textwidth]{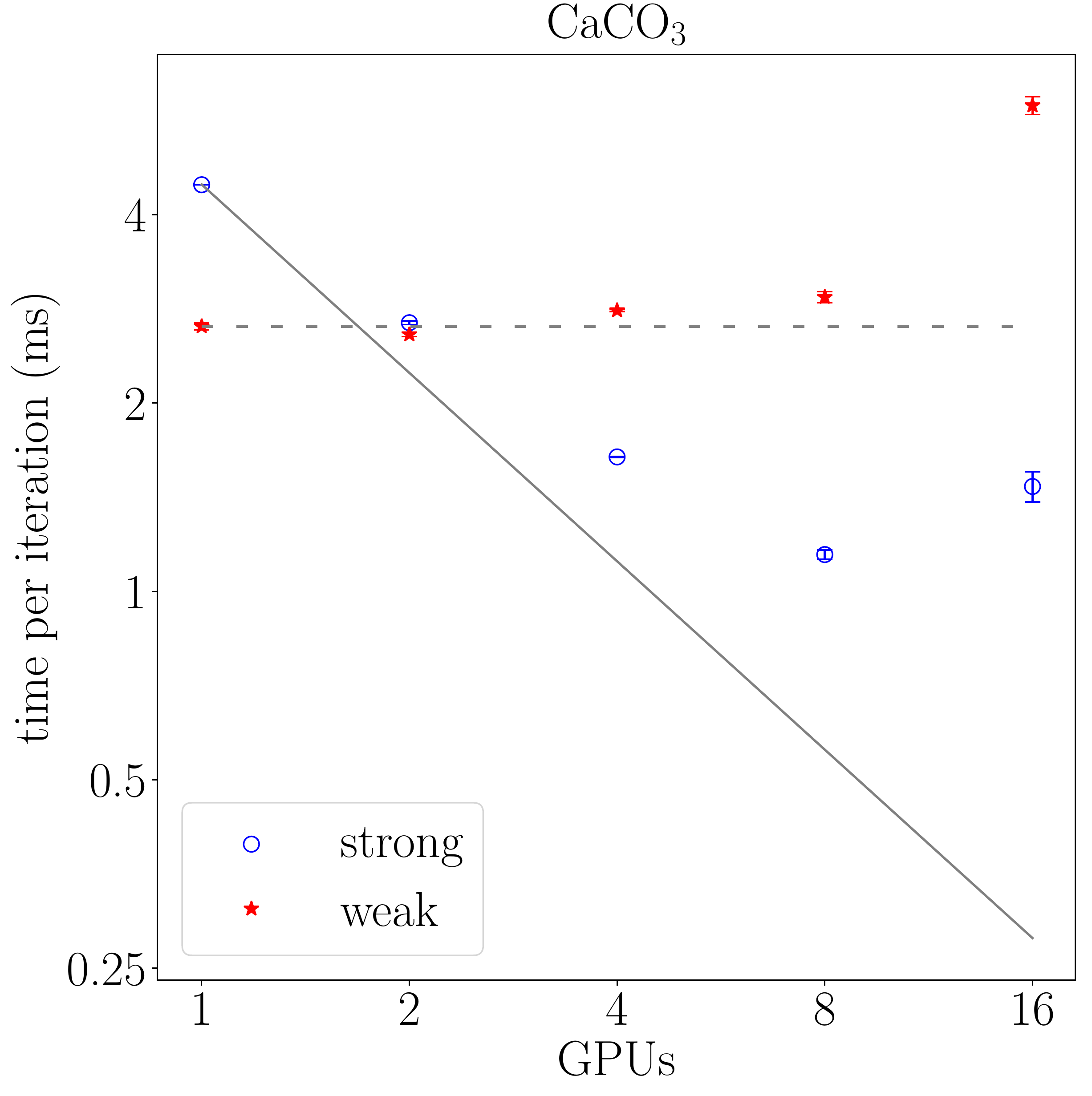}}
\begin{tabular}{ |R{0.4cm}||P{2.5cm}|P{2.2cm}|  }
    \hline
    \multicolumn{3}{|c|}{strong scaling} \\
    \hline
    \# & grid size & time/step (ms) \\
    \hline
    1 & $531 \times 531 \times 531$ & 4.466\,$\pm$\,0.000 \\
    2 & $531 \times 531 \times 531$ & 2.688\,$\pm$\,0.018  \\
    4 & $531 \times 531 \times 531$ & 1.641\,$\pm$\,0.003  \\
    8 & $531 \times 531 \times 531$ & 1.145\,$\pm$\,0.020  \\
    16 & $531 \times 531 \times 531$ & 1.471\,$\pm$\,0.081  \\
    \hline
    \hline
    \multicolumn{3}{|c|}{weak scaling} \\
    \hline
    \# & grid size & time/step (ms) \\
    \hline
    1 & $266 \times 531 \times 531$ & 2.652\,$\pm$\,0.032 \\
    2 & $531 \times 531 \times 531$ & 2.574\,$\pm$\,0.020 \\
    4 & $1062 \times 531 \times 531$ & 2.812\,$\pm$\,0.012 \\
    8 & $1062 \times 1062 \times 531$ & 2.953\,$\pm$\,0.060 \\
    16 & $1062 \times 1062 \times 1062$ & 5.978\,$\pm$\,0.197 \\
    \hline
\end{tabular}
\caption{\rev{Strong and weak parallel scaling of the present implementation on multiple GPUs (\#), simulating inhomogeneous diffusion in a CaCO$_3$ particle-packed bed. The wall clock times are averaged over GPUs and over 80 time steps after 50 steps of GPU warm-up, which are excluded from the averages. All measurements are performed on Nvidia A100 GPUs using FP32 precision. In the plot, blue circles denote strong and red stars weak scaling. Lines show the ideal scaling (solid: strong, dashed: weak). Error bars in the plot and error values in the table give the standard error.}}
\label{fig:applications:multiGPU_CaCO3}
\end{figure}

\begin{figure}[h]
\footnotesize
\centering
\raisebox{-.5\height}{\includegraphics[width=0.45\textwidth]{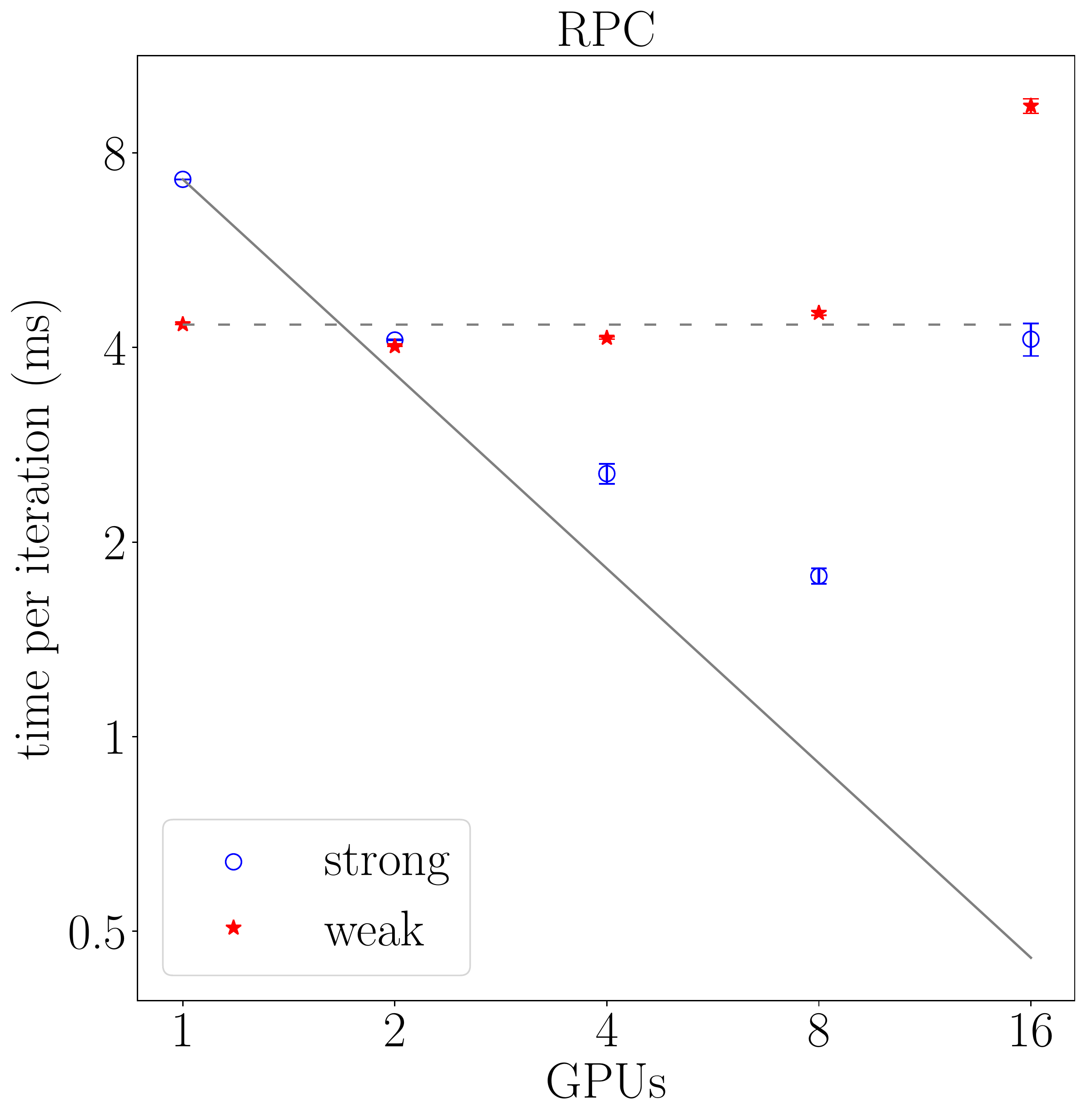}}
\begin{tabular}{ |R{0.4cm}||P{2.5cm}|P{2.2cm}|  }
    \hline
    \multicolumn{3}{|c|}{strong scaling} \\
    \hline
    \# & grid size & time/step (ms) \\
    \hline
    1 & $1216 \times 1016 \times 941$ & 7.280\,$\pm$\,0.000 \\
    2 & $1216 \times 1016 \times 941$ & 4.108\,$\pm$\,0.009 \\
    4 & $1216 \times 1016 \times 941$ & 2.552\,$\pm$\,0.091 \\
    8 & $1216 \times 1016 \times 941$ & 1.772\,$\pm$\,0.049 \\
    16 & $1216 \times 1016 \times 941$ & 4.120\,$\pm$\,0.238 \\
    \hline
    \hline
    \multicolumn{3}{|c|}{weak scaling} \\
    \hline
    \# & grid size & time/step (ms) \\
    \hline
    1 & $608 \times 1016 \times 941$ & 4.343\,$\pm$\,0.001 \\
    2 & $1216 \times 1016 \times 941$ & 4.017\,$\pm$\,0.003 \\
    4 & $2432 \times 1016 \times 941$ & 4.138\,$\pm$\,0.017 \\
    8 & $2432 \times 2032 \times 941$ & 4.522\,$\pm$\,0.038 \\
    16 & $2432 \times 2032 \times 1882$ & 9.458\,$\pm$\,0.245 \\
    \hline
\end{tabular}
\caption{\rev{Strong and weak parallel scaling of the present implementation on multiple GPUs (\#), simulating heat conduction in reticulate porous ceramics (RPC). The wall clock times are averaged over GPUs and over 80 time steps after 50 steps of GPU warm-up, which are excluded from the averages. All measurements are performed on Nvidia A100 GPUs using FP32 precision. In the plot, blue circles denote strong and red stars weak scaling. Lines show the ideal scaling (solid: strong, dashed: weak). Error bars in the plot and error values in the table give the standard error.}}
\label{fig:applications:multiGPU_RPC}
\end{figure}

\rev{We also benchmark the parallel scalability of the code when distributing the problem over an increasing number of GPUs. The results for both weak (problem size increases proportional to GPU count\footnote{\rev{We scale problem size by scaling the size of the background mesh. The number of actually allocated nodes in the sparse grid depends on the geometry, but never deviated more than 1\% from linear scaling.}}) and strong (constant problem size) scaling are shown in Figs.~\ref{fig:applications:multiGPU_CaCO3} (for the CaCO$_3$ case) and \ref{fig:applications:multiGPU_RPC} (for the RPC case). In both figures, the wall-clock times include communication and are averaged over all GPUs and 80 time steps, excluding the first 50 steps of GPU warm-up. For both application cases, good scalability is observed up to 8 GPUs. As expected, weak scaling is better than strong scaling. This is because in strong scaling, the computation time per GPU reduces with increasing GPU count, causing the communication overhead to grow and eventually become limiting. This is exacerbated when using more than a single compute node, i.e.~from 8 to 16 GPUs, requiring communication across nodes via the interconnect network. Taken together, these measurements show that our implementation of the present pipeline scales as expected for different problem sizes and numbers of GPUs.}

\begin{table}[htbp!]
\footnotesize
\centering
\caption{Comparison of memory requirements and wall clock times for one iteration (including all necessary communication) for a dense versus sparse grid implementation of the present pipeline in the two application examples. The sparser the geometry (on the \rev{$8\times 8\times 8$} chunk level), the larger the savings afforded by the sparse grid, as fewer grid chunks are allocated. The wall clock times are averages over \rev{100 time steps per GPU, each after 31 steps of GPU warm-up, which are excluded from the averages. The estimator of the standard deviation of the mean (i.e., the standard error) is given after the $\pm$. All tests are performed on Nvidia A100-SXM4 GPUs using FP32 precision.}
}
\begin{tabular}[t]{ |p{1.4cm}|| P{2.5cm} | R{0.8cm} | R{2cm} | R{0.8cm}| R{2.0cm}|  }
    \hline
    \multirow{3}{*}{\parbox{1.8cm}{case}} & \multirow{3}{*}{\parbox{2.0cm}{grid size}} & \multicolumn{2}{|c|}{dense grid} & \multicolumn{2}{|c|}{sparse grid} \\
    \cline{3-6}
    & & \multicolumn{1}{|c|}{DRAM} & \multicolumn{1}{|c|}{time/step}  & \multicolumn{1}{|c|}{DRAM} & \multicolumn{1}{|c|}{time/step} \\
    & & \multicolumn{1}{|c|}{(GB)} & \multicolumn{1}{|c|}{(ms)}  & \multicolumn{1}{|c|}{(GB)} & \multicolumn{1}{|c|}{(ms)} \\
    
    \hline
    CaCO$_3$    & $531 \times 531 \times 531$       & 2.62  & 5.126\,$\pm$\,0.001    & 2.04 & 4.484\,$\pm$\,0.001 \\
    (1 GPU)     & $266 \times 531 \times 531$       & 1.33  & 2.714\,$\pm$\,0.024    & 1.11 & 2.578\,$\pm$\,0.025 \\
                & $266 \times 266 \times 531$       & 0.67  & 1.487\,$\pm$\,0.016    & 0.59 & 1.507\,$\pm$\,0.000 \\
                & $266 \times 266 \times 266$       & 0.34  & 0.852\,$\pm$\,0.000    & 0.32 & 0.873\,$\pm$\,0.000 \\
    \hline
    RPC         & $1216 \times 1016 \times 941$      & 20.26 & 16.972\,$\pm$\,0.001  & 4.24  & 4.117\,$\pm$\,0.009 \\
    (2 GPUs)    & $608 \times 1016 \times 941$       & 10.26 & 8.739\,$\pm$\,0.001   & 2.54  & 2.622\,$\pm$\,0.006 \\
                & $608 \times 508 \times 941$        & 5.13  & 4.422\,$\pm$\,0.003   & 1.48 & 1.765\,$\pm$\,0.001 \\
                & $608 \times 508 \times 471$        & 2.57  & 2.355\,$\pm$\,0.002   & 0.84 & 1.028\,$\pm$\,0.003 \\ 
    \hline
\end{tabular}
\label{tab:applications:comparison_dense_versus_sparse}
\end{table}

\section{Discussion and conclusions}\label{sec:conclusion}
We have presented a GPU-accelerated simulation pipeline for solving inhomogeneous reaction-diffusion PDEs in real-world, image-based geometries of porous media. Quantitatively understanding the pore-scale dynamics of reaction-diffusion processes is relevant to applications where homogenization or volume-averaging would not be informative. We have demonstrated this \rev{by direct numerical simulation of} fertilizer diffusion in soil and heat conduction in reticulate porous ceramics. In both examples, pore-scale processes such as micropore adhesion or surface heat dissipation could easily and accurately be included in the simulation. The geometry-adapted sparse grid level-set discretization used in the presented pipeline was able to efficiently and effectively cope with the wide spectrum of scales present in fully resolved geometries of porous media.

The implementation of the present pipeline included extending the GPU implementation of distributed sparse block grids~\cite{Incardona2021} to incorporate a level-set method for image-based modeling. The resulting workflow takes microscopy or $\upmu$CT images as input, from which it generates realistic 3D geometries and represents them as a sparse level-set. On the process modeling side, we extended the capabilities of the scalable OpenFPM framework for scientific computing to include space-dependent diffusion coefficients and locally varying reactions. Together, these contributions enabled fully resolved simulations of inhomogeneous reaction-diffusion processes in non-parametric, image-based geometries of porous media.  

After benchmarking the correctness and convergence of the implementation, we showcased the pipeline in two real-world examples. In the first example, we numerically solved an inhomogeneous reaction-diffusion PDE in a sparse grid that discretized the fluid phase of a CaCO$_3$ particle-packed bed. For this example, the sparse grid reduced the memory requirement by 22\% and the runtime by \rev{13\%} compared to a dense grid. In the second example, we simulated heat \rev{conduction} with distributed surface dissipation in the solid phase of reticulate porous ceramics. Reticulate porous ceramics are extremely sparse, and the gain from using a sparse block grid was, therefore, higher than for the CaCO$_3$ particle-packed bed. In particular, the memory requirement was reduced \rev{4.8-fold}, and the simulation runtime was reduced by \rev{76\%}.

\rev{For both application examples, we also benchmarked how the computational cost scales with problem size and with the number of GPUs used. We showed that the present implementation scales approximately linearly with problem size on a single GPU (CaCO$_3$) and on two GPUs (RPC). Parallelizing over more GPUs, we found almost perfect weak scaling and very good strong scaling when using up to 8 GPUs on the same compute node. As expected, the communication overhead became limiting when distributing the simulation across multiple nodes, as we showed for 16 GPUs distributed across 2 compute nodes.}

In macroscopic homogenized models, the impact of a porous geometry on diffusive transport processes can be accounted for by material-dependent constants like tortuosity and effective diffusivity. These constants are thus of great interest, but obtaining them experimentally is cumbersome~\cite{Li2010, Latrille2011, Nguyen2020}. We have therefore shown how the present image-based simulation pipeline enables determining effective diffusion coefficients and tortuosity numerically by simulating FRAP in the transport phase of the diffusion space.

While most currently available image-based simulation pipelines for transport processes in porous media involve commercial software~\cite{An2017}, such as Avizo~\cite{avizo}, Ansys~\cite{ansys}, or COMSOL~\cite{comsol}, the pipeline presented here is entirely free and open-source. The code is written in C++ using the OpenFPM library.

A current limitation of the present pipeline is the accuracy of the boundary approximation, which is first-order. \rev{This is not an intrinsic limitation of the level-set approach, though, which would allow for higher-order boundary treatment by accounting for surface normals and/or curvature. Our approach, however, only requires regular computations on a sparse grid, which is conducive to GPU acceleration as was the main goal here. While other methods for imposing boundary conditions, like immersed boundary and immersed interface methods, require transport phases on both sides of the interface~\cite{Tauriello2015}, this is not the case for the present inhomogeneous diffusion model, where the interface is the actual domain boundary. Finally, penalization methods, which modify the force terms in the momentum equation to account for boundary effects, only apply to flow problems~\cite{Liu2007}. Therefore, we use the SDF to detect the boundary location and impose boundary conditions within the finite-difference stencil, which is well suited for solving reaction-diffusion PDEs, as it is mass-conserving and allows for reactive boundaries. Furthermore, the SDF can directly be used to represent inhomogeneous diffusion coefficients, as shown here.} 

\rev{Obtaining an SDF by level-set redistancing, however, can cause numerical dissipation of the interface geometry. We largely avoided this by using Sussman redistancing, which is interface-preserving~\cite{Sussman1994a, Sussman1999a} and has a discretization error that converges with the correct order (Fig.~\ref{fig:appendix:figures:convergence_sussman}).}
\rev{Our implementation of Sussman redistancing, however, is currently} limited to a dense grid without GPU acceleration, generating a one-off computational overhead at the beginning of a simulation. \rev{Although this is} not a bottleneck for reactive transport simulations \rev{in static geometries, future work will explore whether Sussman redistancing can directly be performed on a sparse grid with GPU acceleration for simulations in dynamic geometries. 
Addition and deletion of points in a sparse grid on the GPU, however, is not efficient, as it requires semaphores to avoid race conditions as well as restructuring or reallocation of memory. Another advantage of performing Sussman redistancing on the host is that its memory capacity is not limited to the DRAM of the GPU. Moreover, our implementation of Sussman redistancing is parallelized for multi-CPU, enabling the computation of SDFs in large grids that need not fit the memory of a single compute node.}

Finally, the current parallel implementation of the present pipeline is limited to explicit time-integration methods. Benefiting from the improved numerical stability of implicit schemes, which allow for larger time-step sizes, would require addressing their larger communication overhead.

Future work will also include applying the present pipeline to other kinds of porous media. In particular, biological tissues could be good candidates, as they tend to have irregular shapes and are porous (e.g., the mineral matrix of bone \cite{Fernandez-Seara2002}, fibrocartilaginous structures \cite{Trampel2002} like meniscus \cite{Travascio2020}, bile-canalicular networks in the liver \cite{Meyer2017,Ahmed2020}, and the extracellular matrix between cells \cite{Wang2016}). In all of these porous media, reaction-diffusion processes are essential to transport water, nutrients, metabolic products, and molecules with regulatory or signaling function. In addition, spatial tissue patterning by graded concentration fields during embryonic development is controlled by a complex interplay between geometry and reaction-diffusion processes~\cite{Wartlick2009, Muller2013Nodal, Muller2013, Umulis2012, Oates2012, Zhang2019}.  
The increasing availability of high-resolution 3D microscopy images of biological tissues provides a unique chance for understanding the organizational principles of morphogenesis~\cite{Kicheva2012, Muller2013, Multerer2018}. We, therefore, anticipate that the generic and efficient computational pipeline presented here will also find application in this field.

\section*{Data availability}\label{sec:data_availability}
The software is available under the GNU General Public License
3.0 (GPLv3) and free of charge from \url{git.mpi-cbg.de/mosaic/software/parallel-computing/openfpm}. We provide the code of this paper on \url{https://git.mpi-cbg.de/mosaic/reactiondiffusion_imagebased_porousmedia.git}. 

\section*{Acknowledgements}\label{sec:acknowledgements}
We thank Prof.~J\"{o}rg Petrasch (Michigan State University, College of Engineering) for providing the $\upmu$CTs of the CaCO$_3$ and RPC samples. We thank Dr.~Pietro Incardona (University of Bonn) for many helpful discussions, proofreading, implementation support with OpenFPM, and the idea of the boundary treatment. From the Sbalzarini group, we thank Dr.~Nandu Gopan, Dr.~Johannes Pahlke, Alejandra Foggia, and Dr.~Aryaman Gupta for discussions and proofreading, and Serhii Yaskovets for support with OpenFPM. We thank Dr.~Michele Marass (MPI-CBG, Dresden) for editorial advice. We thank Dr.~Quentin Vagne (University of Geneva) for helpful discussions. We are grateful to the Scientific Computing Facility of MPI-CBG Dresden and the Center for Information Services and High-Performance Computing (ZIH) of TU Dresden for providing their facilities for the benchmarks. \rev{We thank the anonymous reviewers for their time and constructive suggestions.}

\appendix
\setcounter{section}{0}
\setcounter{figure}{0}   

\section{Additional verification results}\label{sec:appendix}

We provide here below additional information about the analytical solution used in the convergence tests, as well as about the convergence of our implementation of Sussman redistancing of the level-set function.

\subsection{Exact solution for diffusion inside a 2D disk using the method of manufactured solutions}\label{sec:appendix:manufactured_solution}

We derive the exact solution of the test case used to verify our pipeline in Section~\ref{sec:results:convergence}. For this, we use the method of manufactured solutions~\cite{Roache2002}.
The test case considers homogeneous diffusion of a scalar field $u$ \rev{inside} a 2D unit disk with no-flux Neumann boundary conditions and diffusion coefficient $D = 1$. 

A general solution \rev{of Eq.~\eqref{eq:manufactured_solution:homogeneous_diffusion}} proposed in the literature for this case is~\cite{Isaacson2006}:
\begin{equation}{\label{eq:appendix:manufactured_solution:solution}}
    U(r, t) = \left( \frac{r^3}{3} - \frac{r^4}{4} \right) \rm{e}^{-\it B t}.
\end{equation}
We verify that this fulfills the boundary condition:
\begin{equation}{\label{eq:appendix:manufactured_solution:boundary_condition_fulfilled}}
    \left. \frac{\partial U(r, t)}{\partial \bm n} \right\rvert_{r=1}= \frac{\partial U(R, t)}{\partial r} = (R^2-R^3)\rm{e}^{-\it B \it t}=0\, .
\end{equation}
In order to find the particular solution that also fulfills the initial condition, the method of manufactured solutions adds a reaction term $f(r, t)$~\cite{Roache2002}. For this, we write Eq.~\eqref{eq:manufactured_solution:homogeneous_diffusion} as an operator
\begin{equation}{\label{eq:appendix:manufactured_solution:diffusion_operator}}
    L(u) =  
    \pdv{u}{t}
    - \pdv[2]{u}{r} 
    - \frac{1}{r} \pdv{u}{r} 
    - \frac{1}{r^2} \pdv[2]{u}{\theta} 
    = 0.
\end{equation}
and find $f(r, t)$ by applying the operator to $U(r, t)$:
\begin{equation}{\label{eq:appendix:manufactured_solution:source_term}}
    \begin{aligned}
        f(r, t) & = L(U) \\
        & = \pdv{U}{t}
        - \pdv[2]{U}{r} 
        - \frac{1}{r} \pdv{U}{r} 
        - \frac{1}{r^2} \pdv[2]{U}{\theta} \\
        & = \left(\frac{B}{4}r^4 - \frac{B}{3} r^3 + 4 r^2 - 3 r \right) \rm{e}^{-\it B \it t}.
    \end{aligned}
\end{equation}
Using this reaction term in the governing equation, the problem has the exact solution given in Eq.~\eqref{eq:appendix:manufactured_solution:solution}. \rev{The scalar constant $B$ can be arbitrarily chosen and controls the gradient steepness of the initial concentration field. We choose $B=20$, which provides an initial concentration field with smooth gradients and values suitable to be dealt with by single-precision arithmetics.}

\subsection{Convergence of Sussman redistancing}\label{sec:appendix:figures}

\begin{figure}[htbp!]
    \centering
    \includegraphics[width=0.6\textwidth]{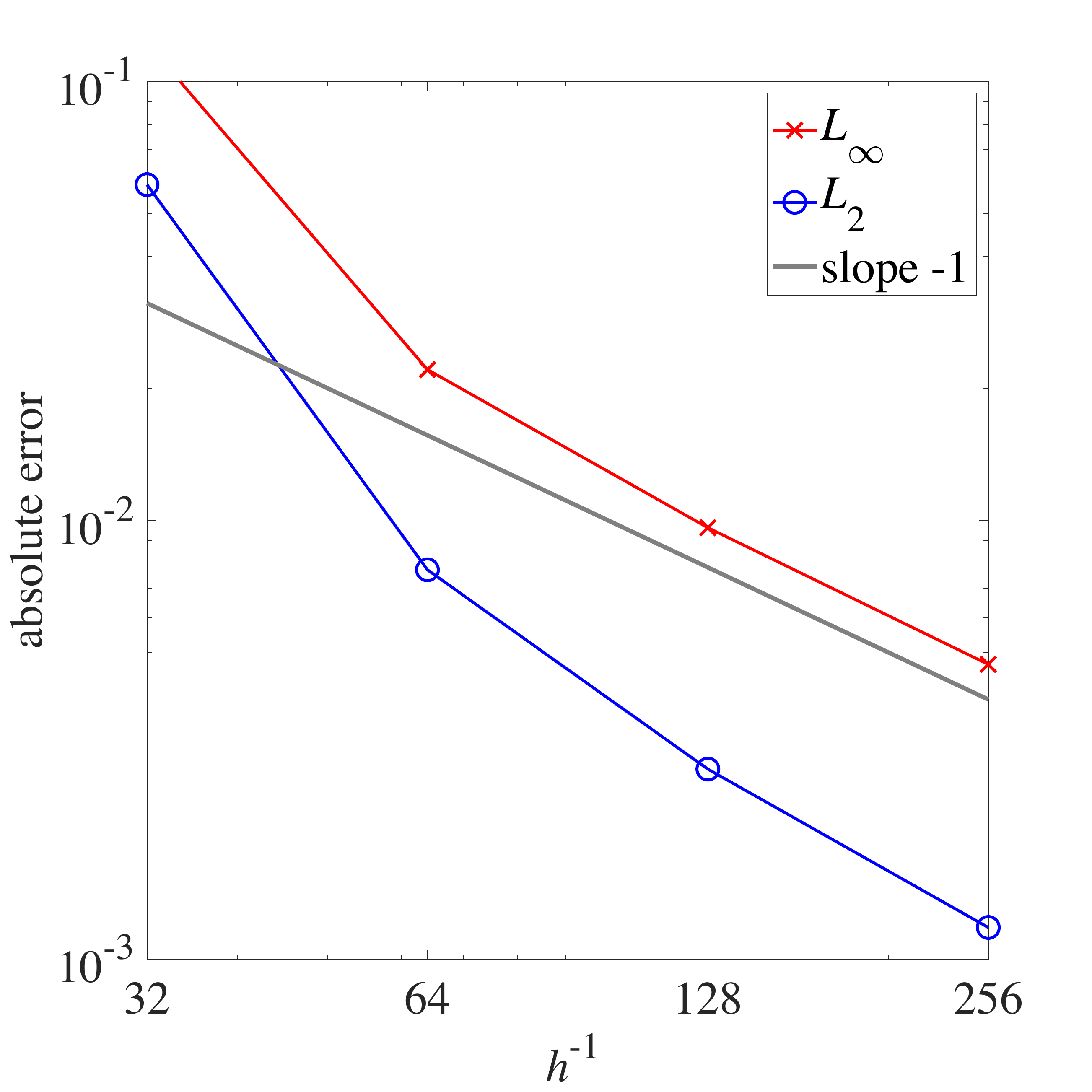}
    \caption{Convergence plot of Sussman redistancing for the level-set of the unit ball in 3D. The $L_2$ (red crosses) and $L_\infty$ (blue circles) norms of the absolute error in the SDF across the entire narrow 
    band of width $4h$ are plotted. Both show the expected linear convergence with the grid spacing $h$ for a fixed time-step size. The upwind gradient computed during iterative reinitialization of the level-set function is approximated using first-order finite differences.}
    \label{fig:appendix:figures:convergence_sussman}
\end{figure}

\rev{In Fig.~\ref{fig:appendix:figures:convergence_sussman}, we present a convergence plot of our implementation of Sussman redistancing of the level-set SDF of the unit ball in 3D using first-order upwind finite differences. The expected order of convergence is achieved as soon as the geometry is well resolved.}

\FloatBarrier

\section{Code examples}\label{sec:appendix:code_examples}

\rev{We provide code examples illustrating the benefits of the OpenFPM-based implementation of our simulation pipeline.
The complete source code of the present implementation is available under the GNU General Public License 3.0 (GPLv3) at~\url{https://git.mpi-cbg.de/mosaic/software/parallel-computing/openfpm} and~\url{https://git.mpi-cbg.de/mosaic/reactiondiffusion_imagebased_porousmedia.git} with examples and documentation at~\url{http://ppmcore.mpi-cbg.de/doxygen/openfpm/index.html}. For demonstration, we only show a few code examples here.}

\rev{To construct a regular, dense grid in OpenFPM, the grid and domain size, ghost layer thickness, and the data-types and sizes of the grid properties have to be defined first. If the sparse will be created based on this dense grid, also the decomposition for the distribution has to be defined. Then, the grid can be allocated.}

\begin{lstlisting}[language=C++,basicstyle=\footnotesize\ttfamily, caption=Distributed dense Cartesian grid construction.]
// Grid type with property aggregate, 
// here just one FP32 that will carry the level-set function
typedef aggregate<float> props;
// Decomposition for distribution
typedef CartDecomposition<dims,float, CudaMemory, 
memory_traits_inte, BoxDistribution<dims,float> > Dec;
typedef grid_dist_id<dims, float, props, Dec > grid_in_type;
// Instantiate distributed regular grid
grid_in_type g_dist(sz, box, ghost);
\end{lstlisting}\label{lst:define_reg_grid}

\rev{The indicator function obtained from the segmentation mask is loaded onto the grid and reinitialized as a SDF using Sussman redistancing, which we implemented as a class template in OpenFPM. Code examples and documentation for this implementation are available at~\url{http://ppmcore.mpi-cbg.de/doxygen/openfpm/example_sussman_images_2D.html}. Here, we  show a snippet of the constructor and how to use it. Iterative redistancing is performed on a temporary grid allocated inside the class with the same domain decomposition as the input grid. Therefore, the fields needed during the redistancing to store the intermediate level-set function, $\phi_n$, its upwind gradient, $\grad \phi_n$, and the sign of the initial $\phi$, $\textrm{sign}(\phi_0)$, as well as the required ghost layers for inter-process communication, are hidden from the user. In our application, the input grid has only one property, namely the pre-redistancing level-set function.}

\begin{lstlisting}[language=C++,basicstyle=\footnotesize\ttfamily, caption=Sussman redistancing class snippet.]
template <typename grid_in_type, 
typename phi_type=double>
class RedistancingSussman
{
public:
	RedistancingSussman(grid_in_type & grid_in, 
	Redist_options<phi_type> &redistOptions) : 
	redistOptions(redistOptions),
	r_grid_in(grid_in), 
	g_temp(grid_in.getDecomposition(),
	grid_in.getGridInfoVoid().getSize(), 
	Ghost<grid_in_type::dims, long int>(3))
	...

\end{lstlisting}\label{lst:sussman_class}

\begin{lstlisting}[language=C++,basicstyle=\footnotesize\ttfamily, caption=Loading the indicator function from a file onto the grid.]
// Initialize grid with (image-based) indicator function
load_pixel_onto_grid<PHI_FULL>(g_dist, 
path_to_zstack, stack_dims);
\end{lstlisting}\label{lst:load_indiactor}

\rev{For running Sussman redistancing, options such as the maximum number of iterations and the convergence tolerance can be passed in the structure {\tt{Redist\_options}} during class instantiation. Redistancing is then executed by calling the method {\tt{run\_redistancing}} with the indices of the input and output properties as template arguments.}

\begin{lstlisting}[language=C++,basicstyle=\footnotesize\ttfamily, caption=Sussman redistancing instantiation and execution.]
// Instantiation of Sussman-redistancing class
RedistancingSussman<grid_type, float> 
redist_obj(g_dist, redist_options);  
// Run Sussman redistancing
redist_obj.run_redistancing<PHI_IN, PHI_OUT>();
\end{lstlisting}\label{lst:sussman}

\rev{Based on the resulting SDF, the sparse grid can be constructed by inserting points only within the diffusion domain, i.e., for points at which \mbox{{\tt{lower\_bound}} $< \phi_{\text{SDF}} <$ {\tt{upper\_bound}}}. Here, \mbox{{\tt{lower\_bound}}} and \mbox{{\tt{upper\_bound}}} are the minimal and maximal values of $\phi_{\text{SDF}}$ within the diffusion domain, depending on the type of level-set function chosen and the phase considered.}

\begin{lstlisting}[language=C++,basicstyle=\footnotesize\ttfamily, caption=Obtain a sparse grid based on the SDF.]
// Create a sparse grid with four FP32 properties and same 
// decomposition Dec as dense grid
typedef aggregate<float, float, float, float> props_sparse;
typedef sgrid_dist_id_gpu<dims, float, props_sparse, 
CudaMemory, Dec> sparse_grid_type;
sparse_grid_type g_sparse(sz, box, ghost);

// At this stage, the sparse grid is still empty. Now, we can loop 
// over the dense grid and insert points in the sparse grid if they 
// lie inside the diffusion domain
template <typename T>
static bool is_diffusionSpace(const T & phi_sdf, 
const T & lower_bound, const T & upper_bound)
{
	const T EPSILON = std::numeric_limits<T>::epsilon();
	const T _b_low = lower_bound + EPSILON;
	const T _b_up  = upper_bound  - EPSILON;
	return (phi_sdf > _b_low && phi_sdf < _b_up);
}

auto dom = grid.getDomainIterator();
while(dom.isNext())
{
	auto key = dom.get();
	if(is_diffusionSpace(grid.template get<PHI_SDF_FULL>(key), 
	b_low, b_up))
	{
		sparse_grid.template 
		insertFlush<PHI_SDF_SPARSE>(key) = 
		grid.template get<PHI_SDF_FULL>(key);
	}
	++dom;
}
\end{lstlisting}\label{lst:obtain_sparse_grid}

\rev{The computations to be run on the sparse grid on the GPUs are defined as a functor. C++ functors are object-like functions, which OpenFPM can pass to the GPU as CUDA or HIP kernels. We demonstrate this exemplary for the case of inhomogeneous diffusion with explicit time stepping.}

\begin{lstlisting}[language=C++, basicstyle=\footnotesize\ttfamily,caption=Defining the functor for inhomogeneous diffusion on the GPU.]
auto epsilon = std::numeric_limits<float>::epsilon();
auto func_inhomogDiffusion = 
[dx, dy, dz, dt, d_low, epsilon] 
__device__ (
float & u_out,          // field out
float & D_out,          // diffusion coefficient out
float & phi_out,        // sdf of domain out
CpBlockType & u,        // field in
CpBlockType & D,        // diffusion coefficient in
CpBlockType & phi,      // sdf of domain in
auto & block, int offset, int i, int j, int k)
{
	// Stencil
	// Field
	float u_c  = u(i, j, k);
	float u_px = u(i+1, j, k);
	float u_mx = u(i-1, j, k);
	... // and so on for y and z
	
	// Signed distance function
	float phi_c  = phi(i, j, k);
	float phi_px = phi(i+1, j, k);
	float phi_mx = phi(i-1, j, k);
	... // and so on for y and z

	// Diffusion coefficient
	float D_c  = D(i, j, k);
	float D_px = D(i+1, j, k);
	float D_mx = D(i-1, j, k);
	... // and so on for y and z		
	
	// Impose no-flux boundary conditions
	if (phi_px <= d_low + epsilon) 
	{u_px = u_c; D_px = D_c;}
	if (phi_mx <= d_low + epsilon) 
	{u_mx = u_c; D_mx = D_c;}
	... // and so on for y and z
	
	// Interpolate diffusion constants between points
	float D_half_px = (D_c + D_px)/2.0;
	float D_half_mx = (D_c + D_mx)/2.0;
	... // and so on for y and z
	
	// Compute concentration of next time point
	u_out = u_c + dt *
	(1/(dx*dx) * (D_half_px * (u_px - u_c) 
	- D_half_mx * (u_c - u_mx)) +
	... // and so on for y and z);

	// Diffusion coefficient and SDF out=in
	D_out 	= D_c; phi_out = phi_c;
};
\end{lstlisting}\label{lst:functor_inhomogDiffusion}

\rev{To solve the diffusion equation over time, this functor and the sparse grid containing the initial condition are downloaded to the GPUs, where they are executed using the convolution functor {\tt{conv3\_b}} with ghost layers communicated between GPUs in a multi-GPU setting.}

\begin{lstlisting}[language=C++, basicstyle=\footnotesize\ttfamily,caption=GPU convolution functor for sparse grids]
// Copy from host to GPU
g_sparse.template hostToDevice<CONC_N, CONC_NPLUS1, 
DIFFUSION_COEFFICIENT, PHI_PHASE>();
// Ghost layer communication of the signed distance function 
// and the inhomogeneous diffusion coefficient
// The SKIP_LABELLING option is only allowed for static geometries
g_sparse.template ghost_get<DIFFUSION_COEFFICIENT, PHI_PHASE>
(RUN_ON_DEVICE | SKIP_LABELLING);
while(iter <= iterations)
{...
    // Update concentration values in ghost layer
    g_sparse.template ghost_get<CONC_N>
    (RUN_ON_DEVICE | SKIP_LABELLING);
    // Convolve functor with sparse grid on GPU
    g_sparse.template conv3_b<CONC_N, DIFFUSION_COEFFICIENT, 
    PHI, CONC_NPLUS1, DIFFUSION_COEFFICIENT, PHI_PHASE, 1>
    ({0, 0, 0}, 
    {(long int) sz[x]-1, ... // and so on for y and z}, 
    func_inhomogDiffusion);
...}
\end{lstlisting}\label{lst:convolve_functor}

\end{document}